\documentclass[pdflatex,iicol]{sn-jnl}

\newcommand{\code}[1]{\texttt{#1}}
\newcommand{\muM}{\delta h_{M}}

\usepackage[numbers,sort&compress]{natbib}
\makeatletter

\def\bibfont{\reset@font\fontfamily{\rmdefault}\normalsize\selectfont}
\makeatother
\usepackage{graphicx}
\usepackage{amsmath,amssymb,amsfonts}
\usepackage{bm}
\usepackage{booktabs}
\usepackage{xcolor}
\let\orcidlogo\relax
\usepackage{orcidlink}

\newcommand{\rscan}{r$^{2}$SCAN}
\newcommand{\rscanu}{r$^{2}$SCAN+$U$}
\newcommand{\epsO}{\delta\mu_{\mathrm{O_2}}}
\newcommand{\dHf}{\Delta H_{\mathrm{f}}}

\begin{document}

\title[Uncertainty-quantified Hubbard \texorpdfstring{$U$}{U} and energy corrections for \texorpdfstring{\rscanu}{r2SCAN+U}]{Uncertainty-quantified Hubbard \texorpdfstring{$U$}{U} and energy corrections for \texorpdfstring{\rscanu}{r2SCAN+U} oxide thermochemistry}

\author*[1]{\fnm{Chikashi} \sur{Shinagawa}\orcidlink{0000-0002-7607-2655}}\email{shinagawa@preferred.jp}
\author*[1]{\fnm{Kohei} \sur{Shinohara}\orcidlink{0000-0002-5907-2549}}\email{kshinohara@preferred.jp}
\author[1]{\fnm{Jun} \sur{Oshiki}\orcidlink{0009-0000-5819-1944}}
\author[1]{\fnm{Tomoya} \sur{Shiota}\orcidlink{0009-0002-7044-4558}}
\author[1]{\fnm{So} \sur{Takamoto}\orcidlink{0000-0001-5739-6497}}
\author[2]{\fnm{Ju} \sur{Li}\orcidlink{0000-0002-7841-8058}}

\affil[1]{\orgname{Preferred Networks, Inc.},
\orgaddress{\street{Otemachi Bldg.\ 1-6-1 Otemachi}, \city{Chiyoda-ku},
\state{Tokyo}, \postcode{100-0004}, \country{Japan}}}

\affil[2]{\orgdiv{Department of Nuclear Science and Engineering, and Department of Materials Science and Engineering}, \orgname{Massachusetts Institute of Technology},
\orgaddress{\city{Cambridge}, \state{MA}, \postcode{02139}, \country{USA}}}

\abstract{Density functional theory guides materials discovery by predicting reaction energies, including battery voltages.
High-throughput calculations now use the \rscan{} functional, to which a Hubbard $U$ can be added for transition metals.
Accurate formation and reaction enthalpies of their oxides further require empirical energy corrections.
However, such corrections are fitted at a $U$ chosen beforehand, and existing \rscanu{} fits omit them or cover only 3d metals.
Here we infer $U$ and an energy correction for each metal by Bayesian inference from experimental enthalpies and cathode voltages.
The data determine $U$ for twelve 3d, 4d and 5d metals, each with its posterior distribution.
The correction scheme lowers the mean absolute error of 145 fitted enthalpies from 0.084 to 0.027~eV/atom.
For seven cathode voltages outside the fit, it cuts the error fourfold, from 0.48 to 0.11~V.
The scheme can thus make high-throughput \rscanu{} materials discovery more reliable for oxides of these metals.}

\maketitle

\section{Introduction}\label{sec:introduction}

Reaction energies and electrochemical voltages determine how compounds transform and how much energy an electrode stores, and predicting them is central to computational materials discovery~\cite{Jain2013,Jain2016}.
Both are differences of free energies, which are commonly approximated by differences of formation enthalpies~\cite{Zhou2004}, and density functional theory (DFT) computes these for any compound at high throughput~\cite{Hohenberg1964,Kohn1965,Hafner2008}.
Their accuracy improves along the hierarchy of semilocal approximations to exchange and correlation, from the local-density (LDA) to the generalized-gradient (GGA) and meta-GGA approximations~\cite{Kohn1965,Perdew1996,Perdew2001}.
In particular, the strongly constrained and appropriately normed (SCAN) meta-GGA improves markedly on the Perdew--Burke--Ernzerhof (PBE) GGA~\cite{Sun2015,Zhang2018}.
Its regularized and restored form, \rscan{}~\cite{Furness2020}, is now run at high throughput.
The Materials Project, a widely used open database of DFT-computed properties of inorganic compounds~\cite{Jain2013}, has adopted it in its calculation workflow~\cite{Kingsbury2022}, and machine-learning interatomic potentials are trained on \rscan{} calculations~\cite{Kaplan2025,Kuner2025,Shinagawa2026}.

For transition-metal oxides, however, the accuracy of SCAN and \rscan{} can be limited.
Their oxidation enthalpies for most 3d transition-metal oxides deviate from experiment, an error attributed mainly to self-interaction in the open-shell d states~\cite{SaiGautam2018,Long2020,Swathilakshmi2023}.
The GGA suffers from the same problem to a greater degree~\cite{Wang2006,Long2020}.
Hybrid functionals, the next level of approximation, describe the redox energetics of transition-metal compounds more accurately than the GGA~\cite{Chevrier2010}.
However, their computational cost, substantially higher than that of a meta-GGA~\cite{Chevrier2010,Zhang2018}, is prohibitive for high-throughput calculation.
The established approach for the GGA has been GGA+$U$~\cite{Wang2006,Jain2011}, which adds a Hubbard-like on-site interaction for localized electrons, characterized by an effective parameter $U$~\cite{Anisimov1991,Dudarev1998}.
The same approach has been adopted for SCAN and \rscan{}, giving SCAN+$U$ for oxides~\cite{SaiGautam2018,Long2020,Artrith2022} and \rscanu{} for oxides~\cite{Swathilakshmi2023}, fluorides~\cite{Tekliye2024} and sulfides~\cite{Donovan2025}.
The formation enthalpies of these oxides computed with \rscan{} alone nevertheless deviate little from experiment on average, because an error of the O$_2$ molecule partly cancels an error of opposite sign (\nameref{sec:fit-agreement}, below).
Adding a Hubbard $U$ removes only the latter and leaves the error of O$_2$ exposed.

Formation enthalpies computed with a semilocal functional carry systematic errors, and these are commonly reduced by empirical energy corrections fitted to experimental formation enthalpies~\cite{Grindy2013,Friedrich2019,Aykol2014,Gong2022,Bartel2018,Reents2026}.
The fitted elemental-phase reference energies (FERE) scheme replaces the computed energy of each unary phase with a fitted reference energy~\cite{Lany2008,Stevanovic2012}.
Another line of work corrected the overbinding of the O$_2$ molecule with an energy shift~\cite{Wang2006} and introduced a per-metal shift that makes GGA+$U$ energies of transition-metal oxides comparable with GGA energies of the metals~\cite{Jain2011}.
The Materials Project's MP2020 scheme extends this line of work, fitting the O$_2$ correction together with the per-metal shifts, with their uncertainties~\cite{Wang2021}.
In SCAN+$U$ and \rscanu{} as well, the metal is computed without a Hubbard $U$, because its electrons are delocalized~\cite{Artrith2022,Donovan2025}.
Since total energies computed at different values of $U$ cannot be compared directly~\cite{Aykol2014}, a formation enthalpy that combines an \rscanu{} oxide with an \rscan{} metal needs an empirical energy correction such as the per-metal shift.
The Hubbard $U$ and the corrections therefore jointly determine a computed formation enthalpy and must be chosen consistently.
In both schemes, however, the corrections are fitted at values of $U$ chosen beforehand.
FERE fixes $U$ for the transition metals before fitting its reference energies~\cite{Stevanovic2012}, and the MP2020 scheme refits its corrections at fixed $U$~\cite{Wang2021}.

The Hubbard $U$ can be computed from first principles for each site of each structure, by the linear response approach~\cite{Cococcioni2005,Zhou2004,Kulik2011,Moore2024,Bastonero2025} or the constrained random-phase approximation (cRPA)~\cite{Aryasetiawan2004}.
It can also be chosen to reproduce electronic properties, such as band gaps, band structures and magnetic moments~\cite{Yu2020,Cai2024,Tavadze2021}.
Machine-learning models trained on values of either kind now predict a structure-dependent $U$ for specific systems, namely Fe--O~\cite{Xia2024}, Mn--O~\cite{Cai2024,Cao2026}, and Mn, Fe and Ni oxides and phosphates for lithium-ion cathodes~\cite{Uhrin2025}.
A $U$ computed separately for each structure changes the Hamiltonian from one configuration to the next~\cite{Warford2026}.
Finally, $U$ can be fitted to experimental formation or reaction enthalpies, as in GGA+$U$~\cite{Wang2006,Aykol2014}, SCAN+$U$~\cite{SaiGautam2018,Long2020,Artrith2022} and \rscanu{}~\cite{Swathilakshmi2023,Donovan2025}.
In this work, we focus on thermochemistry and therefore take the last approach, fitting $U$ to experimental formation enthalpies, reaction enthalpies and lithium-ion cathode voltages.
A formation enthalpy and a cathode voltage are both differences of energy between compounds in which the transition metal takes different oxidation states, and the two therefore constrain the same $U$.
We fit one $U$ per element, because we have in mind high-throughput calculations~\cite{Jain2013,Jain2016,Oganov2019,Merchant2023,Horton2025}, in particular the training datasets of universal machine-learning interatomic potentials, which require a smooth potential energy surface.
Existing \rscanu{} values of $U$ for oxides were fitted without an O$_2$ correction or per-metal shift~\cite{Swathilakshmi2023}, and such corrections could further improve the formation enthalpies computed with them.
The SCAN+$U$ fit of oxides~\cite{Artrith2022} and the \rscanu{} fit of sulfides~\cite{Donovan2025} that include a per-metal shift cover only the 3d metals.

A further difficulty is that each value of $U$ rests on a few experimental values, which can constrain it only weakly.
One value per element cannot follow the variation of $U$ between compounds~\cite{Moore2024}, the fitted $U$ is correlated with the per-metal shift, and formation enthalpies and voltages can favour different values.
Existing fits address this only in part.
The GGA+$U$ fit behind the Materials Project's values of $U$ rests on the experimental enthalpies of two to four oxides per metal~\cite{Wang2006,Jain2011}, and the joint SCAN+$U$ fit reports the standard deviation of each parameter over leave-one-out cross-validation, a measure of its sensitivity to the choice of reference data~\cite{Artrith2022}.
We therefore determine each value by Bayesian inference~\cite{Murphy2012} rather than as a point estimate, and the width of its posterior distribution shows how well the experimental data constrain it.
Bayesian methods have been applied to $U$ before.
Bayesian optimization returns a single value of $U$ fitted to hybrid-functional band gaps and band structures~\cite{Yu2020,Popov2021,Yu2023}.
Bayesian inference has given a posterior distribution of $U$ from volume, magnetic moment and band gap~\cite{Tavadze2021}, whereas the posterior distribution of this work is inferred from formation enthalpies and lithium-ion cathode voltages.

Here we establish a parameterization of \rscanu{} for oxides of the 3d, 4d and 5d metals in the functional form of the MP2020 scheme.
The O$_2$ correction is fitted first and separately, on oxides to which no Hubbard $U$ applies.
The values of $U$ and the per-metal shifts are then inferred together, against experimental formation enthalpies and lithium-ion cathode voltages, and each value of $U$ is thus determined within the correction scheme and carries a posterior distribution.

\section{Results}\label{sec:results}

\newcommand{\FitEnthalpies}{145}
\newcommand{\FitVoltages}{11}
\newcommand{\PreliminaryEnthalpies}{172}
\newcommand{\PreliminaryMetals}{16}
\newcommand{\EnthalpiesNoMetal}{13}
\newcommand{\FitDataRows}{%
Ti & 23 & 2 \\
V & 16 & 1 \\
Cr & 22 & 0 \\
Mn & 13 & 3 \\
Fe & 20 & 1 \\
Co & 9 & 3 \\
Ni & 7 & 1 \\
Cu$^{\dagger}$ & 8 & 0 \\
Nb$^{\dagger}$ & 6 & 0 \\
Mo & 22 & 0 \\
Tc & 3 & 0 \\
Ru & 2 & 0 \\
Ag$^{\dagger}$ & 7 & 0 \\
W & 19 & 0 \\
Re$^{\dagger}$ & 7 & 0 \\
Os & 2 & 0 \\
}
\newcommand{\OtwoValues}{52}
\newcommand{\OtwoBinaryValues}{21}
\newcommand{\OtwoMulticationValues}{31}
\newcommand{\OtwoEps}{0.31}
\newcommand{\OtwoMeanDifference}{0.31}
\newcommand{\OtwoMinDifference}{0.10}
\newcommand{\OtwoMaxDifference}{0.63}
\newcommand{\OtwoStdDifference}{0.11}
\newcommand{\OtwoMAEUncorrected}{0.088}
\newcommand{\OtwoMAECorrected}{0.022}
\newcommand{\ULowCo}{0.34}
\newcommand{\UFe}{4.7}
\newcommand{\ULowFe}{3.09}
\newcommand{\UHighFe}{6.23}
\newcommand{\ULowTi}{0.07}
\newcommand{\ULowW}{0.11}
\newcommand{\TauEnthalpy}{40}
\newcommand{\MeanSigmaEnthalpy}{9.5}
\newcommand{\TauVoltage}{0.21}
\newcommand{\MeanSigmaVoltage}{0.084}
\newcommand{\UNoVoltageFe}{4.5}
\newcommand{\FittedParameterRows}{%
Ti & 1.4 & 0.07--2.39 & $-0.587 \pm 0.036$ \\
V & 1.6 & 1.28--1.84 & $-0.434 \pm 0.035$ \\
Cr & 1.0 & 0.71--1.23 & $-0.125 \pm 0.037$ \\
Mn & 1.7 & 1.34--2.10 & $-0.214 \pm 0.038$ \\
Fe & 4.7 & 3.09--6.23 & $-1.423 \pm 0.033$ \\
Co & 2.3 & 0.34--2.93 & $-0.720 \pm 0.049$ \\
Ni & 2.4 & 1.08--3.78 & $-0.694 \pm 0.060$ \\
Mo & 2.7 & 2.09--3.23 & $-1.867 \pm 0.046$ \\
Tc & 1.5 & 0.52--2.42 & $-0.497 \pm 0.111$ \\
Ru & 2.6 & 1.41--3.78 & $-1.474 \pm 0.099$ \\
W & 3.0 & 0.11--4.39 & $-2.297 \pm 0.053$ \\
Os & 2.6 & 1.43--3.54 & $-1.828 \pm 0.099$ \\
}
\newcommand{\ScoreEnthalpies}{145}
\newcommand{\ScoreMAEEnthalpyRscan}{0.084}
\newcommand{\ScoreMAEEnthalpyScheme}{0.027}
\newcommand{\ScoreMeanRscan}{-0.050}
\newcommand{\ScoreMeanEps}{-0.142}
\newcommand{\ScoreMeanOtwo}{0.092}
\newcommand{\ScoreMeanScheme}{+0.001}
\newcommand{\ScoreVoltages}{11}
\newcommand{\ScoreMAEVoltageRscan}{0.30}
\newcommand{\ScoreMAEVoltageScheme}{0.17}
\newcommand{\ScoreMeanVoltageRscan}{-0.30}
\newcommand{\ScoreMeanVoltageScheme}{-0.09}
\newcommand{\ScoreVoltageImproved}{8}
\newcommand{\OutsideVoltages}{7}
\newcommand{\OutsideMAERscan}{0.48}
\newcommand{\OutsideMAEScheme}{0.11}
\newcommand{\OutsideMAEFactor}{4.4}
\newcommand{\OutsideMAENoVoltage}{0.13}
\newcommand{\OutsideMeanRscan}{-0.48}
\newcommand{\OutsideMeanScheme}{-0.08}

\subsection{Design of the correction scheme}\label{sec:design}

\subsubsection{The correction scheme}\label{sec:correction-scheme}

In this work, we correct the total energies of oxides, meaning compounds in which O is an anion, including those that also contain another anion such as F.
Compounds that contain no O as an anion, fluorides among them, are not treated.
The correction depends only on composition, as in the MP2020 scheme and the corrections it builds on~\cite{Wang2006,Jain2011,Wang2021}.
The formation enthalpy of an oxide $M_{n_M} M'_{n_{M'}} \cdots \mathrm{O}_{n_{\mathrm{O}}}$ calculated without correction is
\begin{equation}
\dHf^{\mathrm{calc}}(U) = \frac{1}{N} \Bigl( E(U) - \sum_{M} n_M \mu_M - \frac{n_{\mathrm{O}}}{2} \mu_{\mathrm{O_2}} \Bigr) ,
\label{eq:uncorrected-formation-enthalpy}
\end{equation}
where $N = \sum_M n_M + n_{\mathrm{O}}$ is the number of atoms in the formula, $E(U)$ is the calculated total energy of the formula, with the Hubbard $U_M$ on each metal $M$ for which this work determines one and without $U$ on any other metal, the argument $U$ collects these $U_M$, and the chemical potentials $\mu_M$ and $\mu_{\mathrm{O_2}}$ are the calculated energy per atom of the unary phase of $M$ and the calculated energy of the O$_2$ molecule.
The index $M$ runs over every element of the formula other than O, including any anion other than O.
The unary phases are computed without a Hubbard $U$ (\nameref{sec:methods-unary-phases}).
With every $U_M$ set to zero, $\dHf^{\mathrm{calc}}(0)$ is the value of \rscan{} alone.
The argument is omitted for an oxide in which no metal takes a Hubbard $U$.
The corrected formation enthalpy is
\begin{equation}
\begin{aligned}
\dHf^{\mathrm{corr}}(U, \delta h) ={}& \dHf^{\mathrm{calc}}(U) \\
&+ \frac{1}{N} \Bigl( \sum_{M} n_M \muM - \frac{n_{\mathrm{O}}}{2} \epsO \Bigr) .
\end{aligned}
\label{eq:corrected-formation-enthalpy}
\end{equation}
The O$_2$ correction $\epsO$ shifts the chemical potential of O$_2$ and thus corrects the binding energy of the molecule, and applies to every oxide, including those of metals that take no Hubbard $U$.
The per-metal shift $\muM$ is a correction to the formation enthalpy itself, specific to each metal that takes a Hubbard $U$ and zero for the others, and the argument $\delta h$ collects these shifts.
Compounds other than oxides, the unary phases among them, are computed with \rscan{} without a Hubbard $U$ and take no correction, and they enter the same phase diagram as the oxides.
The O$_2$ correction is determined first, independently of $U$, and $U_M$ and $\muM$ are determined afterwards together, in \nameref{sec:fitting-parameters}.

\subsubsection{The O\texorpdfstring{$_2$}{2} correction}\label{sec:o2-correction-results}

The O$_2$ correction is calibrated on the experimental formation enthalpies of oxides to which no Hubbard $U$ applies, because an O$_2$ correction fitted on them cannot absorb the effect of $U$.
These are oxides of main-group elements and of Sc and Zn, whose ions have empty or filled d shells.
Oxides of nine heavy p-block elements (As, Se, In, Sn, Sb, Te, Tl, Pb and Bi) are left out.
The energies of their unary phases are sensitive to spin-orbit coupling and to dispersion interactions, neither of which the calculations here include, and an error in these energies would be absorbed into the O$_2$ correction.
An O$_2$ correction of this form fitted with other functionals depends strongly on the fitting set once p-block oxides are included~\cite{Friedrich2019}.
In this work, the calibration is kept to oxides whose calculated energies can be trusted, and heavy elements, which a reliable calculation would have to treat with spin-orbit coupling at higher cost, are avoided.

\begin{figure}[t]
\centering
\includegraphics[width=\linewidth]{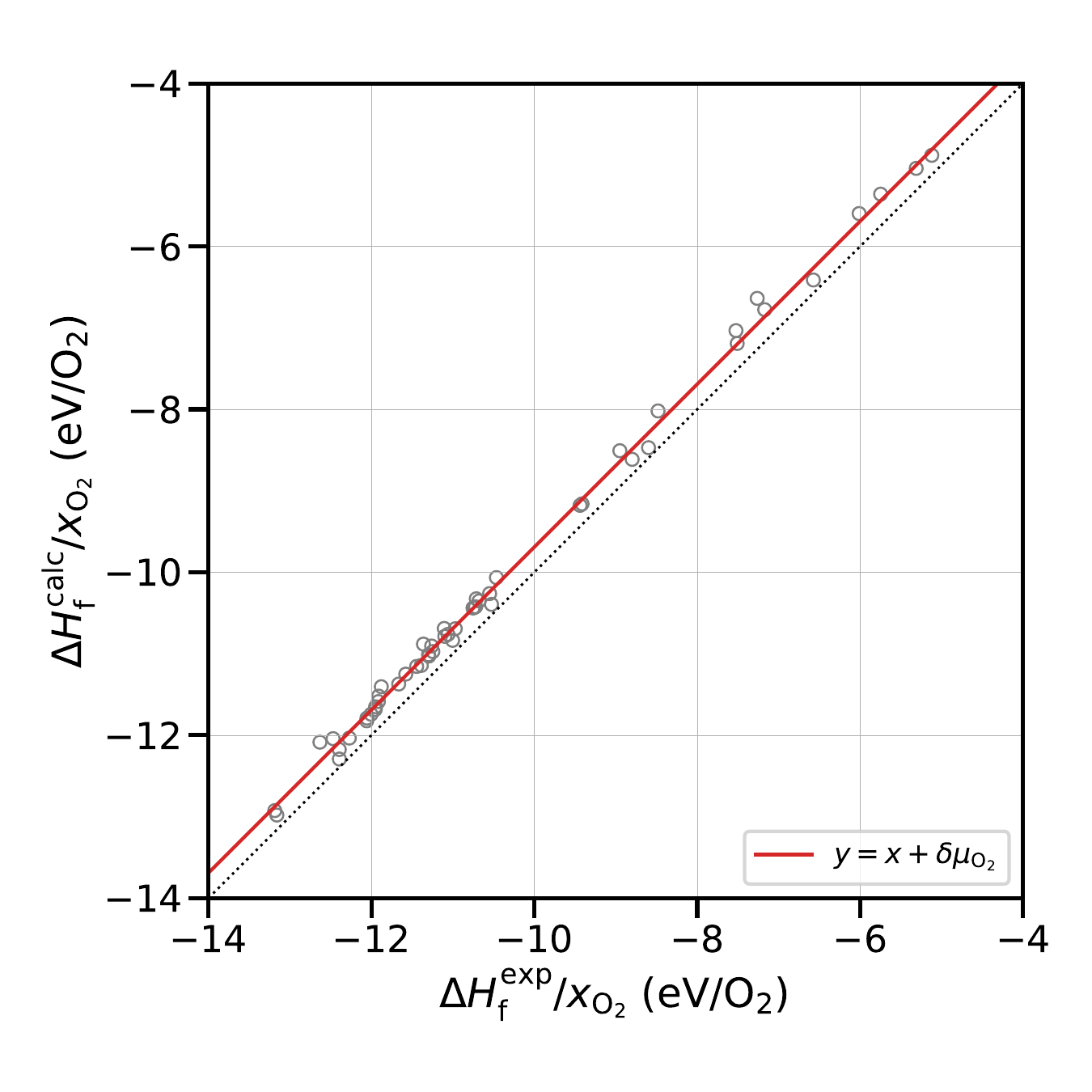}
\caption{\textbf{The O$_2$ correction on the oxides it is calibrated on.}
The \rscan{} formation enthalpy per O$_2$ molecule, $\dHf^{\mathrm{calc}}/x_{\mathrm{O_2}}$, against the experimental one, $\dHf^{\mathrm{exp}}/x_{\mathrm{O_2}}$, for each experimental value the correction is calibrated on, where $x_{\mathrm{O_2}}$ is the number of O$_2$ molecules in the formation reaction per atom of the compound.
The dotted line marks equal enthalpies, and the solid line is offset from it by $\epsO$.}
\label{fig:o2-correction}
\end{figure}

Without the O$_2$ correction, the \rscan{} formation enthalpies of these oxides are less negative than experiment by \OtwoMeanDifference{}~eV per O$_2$ on average (Fig.~\ref{fig:o2-correction}), and the O$_2$ correction is $\epsO = \OtwoEps \pm \OtwoStdDifference$~eV per O$_2$ (\nameref{sec:o2-correction}).
The \rscan{} formation enthalpy is less negative than experiment for every calibration value, by \OtwoMinDifference{} to \OtwoMaxDifference{}~eV per O$_2$.
The correction lowers the mean absolute error on these oxides from \OtwoMAEUncorrected{} to \OtwoMAECorrected{}~eV per atom.

\subsubsection{Experimental and computed data for transition-metal oxides}\label{sec:data}

We consider the 24 metals Ti--Cu, Zr--Ag and Hf--Au.
The lanthanoids and actinoids are left out, because their Hubbard $U$ acts on f rather than d shells and the experimental enthalpies of their oxides are too few to determine it.
The per-metal shift $\muM$ in Eq.~\eqref{eq:corrected-formation-enthalpy} adds the same amount for every atom of a metal and thus absorbs the part of the change with $U$ that is common to all its compounds.
Compounds at one oxidation state change with $U$ at similar rates, and we therefore require experimental enthalpies at two or more oxidation states of a metal to determine its $U$.

In the experimental enthalpies curated in this work, eight of the 24 metals (Zr, Rh, Pd, Hf, Ta, Ir, Pt and Au) have values at one oxidation state or none.
Table~\ref{tab:fit-data} lists the number of experimental values for each of the other \PreliminaryMetals{}.
We first fit these \PreliminaryMetals{} metals together (\ref{si-sn:all-metals}), with $U_{\mathrm{Re}}$ and $\delta h_{\mathrm{Re}}$ held at zero, because the \rscanu{} calculations of Re oxides fail at large $U$.
The data determine $U$ for twelve of them (Ti to Ni, Mo, Tc, Ru, W and Os).
For Nb and Cu, the posterior distributions do not separate a positive $U$ from zero, and for Ag the 95\% interval spans most of the prior range.
The Re enthalpies deviate from the calculated values by similar amounts, which a per-metal shift would absorb and $U_{\mathrm{Re}}$ would not (\nameref{sec:fitted-parameters}).
The values of $U_M$ and $\muM$ reported in this work are those of a second fit to the twelve alone, which reads \FitEnthalpies{} experimental formation and reaction enthalpies of oxides under ambient conditions and \FitVoltages{} lithium-ion cathode voltages.

\begin{table}[t]
\caption{Experimental values for each metal in the first fit. An enthalpy of a compound with two of these metals counts toward both, and another \EnthalpiesNoMetal{} are of oxides that contain none of them. The fit to the twelve leaves out the metals marked $\dagger$ and the enthalpies of their compounds, reading \FitEnthalpies{} of the \PreliminaryEnthalpies{} enthalpies.}
\label{tab:fit-data}
\begin{tabular*}{\columnwidth}{@{\extracolsep\fill}lrr}
\toprule
Metal & Enthalpies & Voltages \\
\midrule
\FitDataRows
\bottomrule
\end{tabular*}
\end{table}

The energy of a transition-metal oxide depends on its magnetic configuration, and the lowest-energy configuration can change with $U$.
Existing SCAN+$U$ and \rscanu{} fits mostly adopt the configuration reported by experiment~\cite{SaiGautam2018,Long2020,Swathilakshmi2023}.
In the high-throughput calculations the scheme is for, however, the configuration is not known in advance and is the lowest-energy one that DFT finds.
In this work, we therefore determine the configuration of every compound by DFT, the protocol under which the fitted values are used (\nameref{sec:methods-computed-data}).
For each compound, the fit reads the DFT energy (\nameref{sec:methods-dft-calculations}) of the crystal structure assigned to it (\nameref{sec:methods-computed-data}) in its lowest-energy magnetic configuration.
The total energies are calculated at 0~K and compared with experimental standard formation enthalpies at 298.15~K, without considering the temperature dependence between the two.
Computing the temperature dependence with \rscanu{} at every value of $U$ would be prohibitively costly.

\subsubsection{Fitting the parameters}\label{sec:fitting-parameters}

With $\epsO$ held fixed, $U_M$ and $\muM$ are inferred together by Bayesian inference against both the formation enthalpies and the cathode voltages (\nameref{sec:fitting-u-mu}).
The voltages constrain $U_M$ alone, because $\muM$ and $\epsO$ cancel from a voltage (\nameref{sec:methods-cathode-voltages}).
Existing fits of $U$ to experimental enthalpies compare enthalpies of oxidation between binary oxides, each the difference of two experimental formation enthalpies~\cite{Wang2006,SaiGautam2018,Long2020,Swathilakshmi2023}, or the formation enthalpies of binary and ternary oxides together with every reaction derived from them~\cite{Artrith2022}.
In this work, each experimental value instead enters the fit once, as the reaction it measures, and the values are treated as independent observations.

Each experimental value is modelled as normally distributed about its corrected calculated value,
\begin{equation}
\begin{aligned}
\dHf^{\mathrm{exp}}(x) &\sim \mathcal{N}\bigl(\dHf^{\mathrm{corr}}(x, U, \delta h),\ s_{\mathrm{f}}(x)^2 + \tau_{\mathrm{H}}^2\bigr) , \\
\Delta H_{\mathrm{c}}^{\mathrm{exp}}(x) &\sim \mathcal{N}\bigl(\Delta H_{\mathrm{c}}^{\mathrm{corr}}(x, U),\ s_{\mathrm{c}}(x)^2 + \tau_{\mathrm{H}}^2\bigr) , \\
V^{\mathrm{exp}}(x) &\sim \mathcal{N}\bigl(V^{\mathrm{corr}}(x, U),\ s_V(x)^2 + \tau_V^2\bigr) ,
\end{aligned}
\label{eq:observation-model}
\end{equation}
for a formation enthalpy (Eq.~\eqref{eq:corrected-formation-enthalpy}), an enthalpy of reaction from binary oxides (Eq.~\eqref{eq:conversion-enthalpy}) and a cathode voltage (Eq.~\eqref{eq:voltage}).
Here $x$ is the crystal structure of the oxide, or the structures of the compounds in the reaction or the cathode couple, $s_{\mathrm{f}}$, $s_{\mathrm{c}}$ and $s_V$ are the uncertainties stated by the source of each value, and $\tau_{\mathrm{H}}$ and $\tau_V$ are model errors, one for the enthalpies and one for the voltages.
The model errors $\tau_{\mathrm{H}}$ and $\tau_V$ are introduced to represent the differences between the calculated and experimental values that no stated uncertainty covers, such as those due to the functional, to the difference between energies at 0~K and enthalpies at 298.15~K, and to systematic errors of the experiments.

The model of Eq.~\eqref{eq:observation-model} assigns a probability to each experimental value for given $x$, $U$, $\delta h$ and $\tau = (\tau_{\mathrm{H}}, \tau_V)$, and the product of these probabilities over the set $\mathcal{Y}$ of all the experimental values and the set $\mathcal{X}$ of their structures is the likelihood $p(\mathcal{Y} \mid \mathcal{X}, U, \delta h, \tau)$.
By Bayes' theorem, the posterior distribution of $U$, $\delta h$ and $\tau$ is proportional to this likelihood multiplied by their prior distributions,
\begin{equation}
\begin{split}
& p(U, \delta h, \tau \mid \mathcal{X}, \mathcal{Y}) \\
& \quad \propto p(U)\, p(\delta h)\, p(\tau)\, p(\mathcal{Y} \mid \mathcal{X}, U, \delta h, \tau) ,
\end{split}
\label{eq:posterior}
\end{equation}
where the priors $p(U)$, $p(\delta h)$ and $p(\tau)$ are given in \nameref{sec:fitting-u-mu}, and the posterior is sampled by Markov chain Monte Carlo~\cite{Murphy2012}.
Sampling requires the energy of each compound at any value of $U$, and it is interpolated between the values at which DFT computes it (\nameref{sec:methods-energy-curves}).

\subsection{Fitted parameters}\label{sec:fitted-parameters}

\begin{figure*}[t]
\centering
\includegraphics[width=\textwidth]{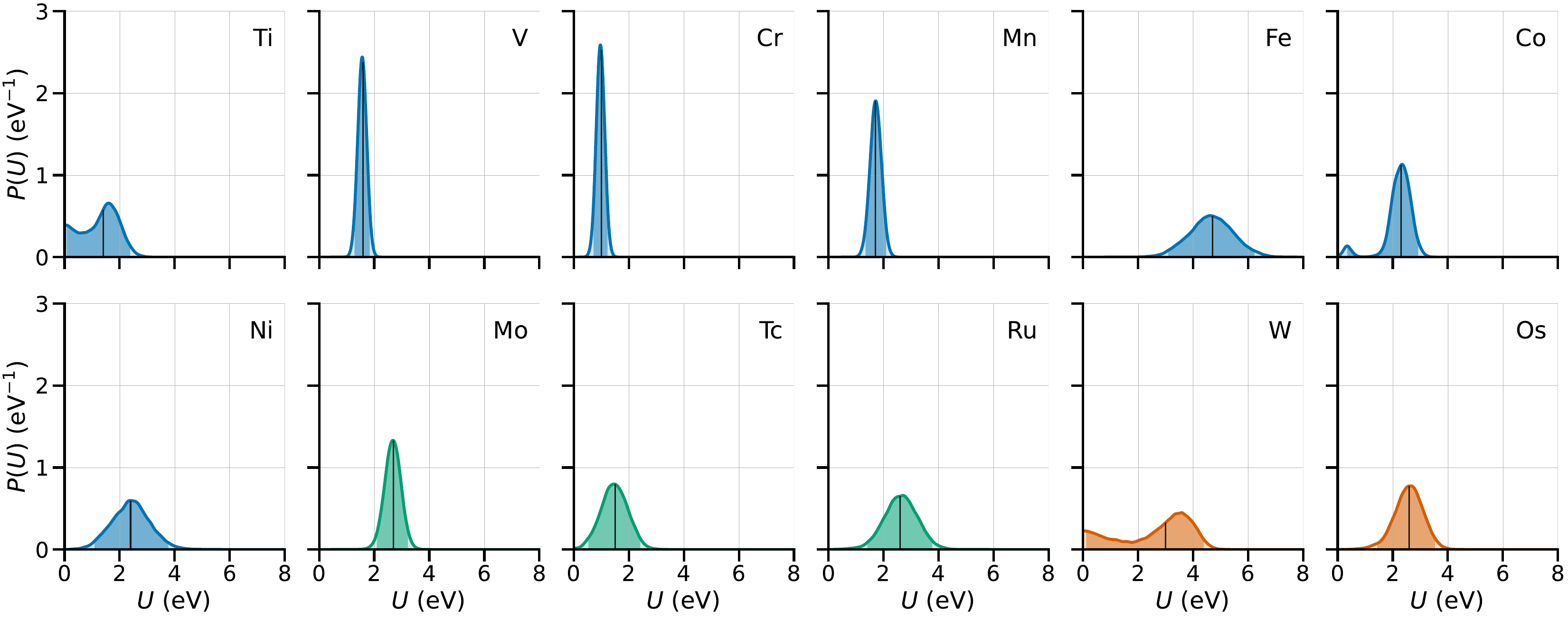}
\caption{\textbf{Marginal posterior distribution of the Hubbard $U$ of each metal.}
Each panel shows the density of the posterior draws of $U_M$ over the range of its prior, with the 95\% interval shaded darker.
The vertical line marks the value of $U_M$ in Table~\ref{tab:fitted-parameters}, and the colour gives the period of the metal.}
\label{fig:posterior-marginal}
\end{figure*}

Table~\ref{tab:fitted-parameters} lists the fitted $U_M$ and $\muM$, and Fig.~\ref{fig:posterior-marginal} shows the marginal posterior distribution of each $U_M$.
We report each $U_M$ as the median of its posterior distribution, rounded to 0.1~eV, with its equal-tailed 95\% interval.
Although each experimental value is normally distributed about its calculated value in Eq.~\eqref{eq:observation-model}, the posterior distribution of $U_M$ is not normal, because the calculated values depend on $U$ nonlinearly through the energies of the compounds.
The marginal distributions of most metals have a single peak, and those of Co, Ti and W have a second well at small $U$, which the interval then spans.
The median is therefore reported rather than the maximum of the posterior.
The per-metal shift $\muM$ is evaluated at the rounded value of $U_M$, because it is valid only for the $U_M$ it is fitted with.

\begin{table}[t]
\caption{Fitted parameters. For each metal, the Hubbard $U$ as the median of its posterior distribution rounded to 0.1~eV, its equal-tailed 95\% interval, and the per-metal shift $\muM$ per atom of $M$ evaluated at the rounded $U$, with its standard deviation.}
\label{tab:fitted-parameters}
\begin{tabular*}{\columnwidth}{@{\extracolsep\fill}lrrr}
\toprule
Metal & $U$ (eV) & 95\% interval (eV) & $\muM$ (eV) \\
\midrule
\FittedParameterRows
\bottomrule
\end{tabular*}
\end{table}

The posterior medians of $\tau_{\mathrm{H}}$ and $\tau_V$ are \TauEnthalpy{}~meV per atom and \TauVoltage{}~V.
The median of $\tau_{\mathrm{H}}$ is about four times the mean stated uncertainty of the enthalpies, \MeanSigmaEnthalpy{}~meV per atom, and that of $\tau_V$ more than twice the mean stated uncertainty of the voltages, \MeanSigmaVoltage{}~V.
The model errors thus dominate the uncertainty with which each experimental value enters the fit.

\begin{figure*}[t]
\centering
\includegraphics[width=\textwidth]{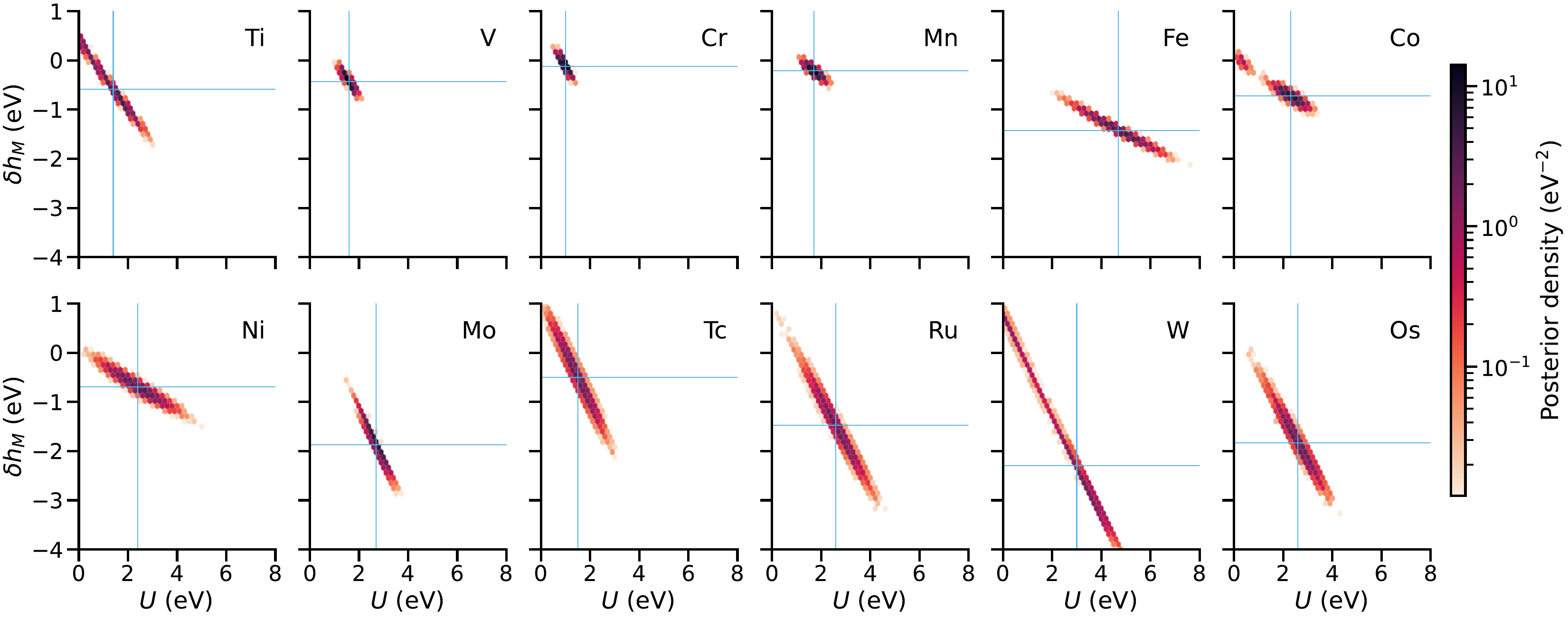}
\caption{\textbf{Joint posterior distribution of the Hubbard $U$ and the per-metal shift $\muM$ of each metal.}
Each panel shows the posterior density of $(U_M, \muM)$ in hexagonal bins of the posterior draws, on one logarithmic colour scale for every panel.
The vertical and horizontal lines mark the values of $U_M$ and $\muM$ in Table~\ref{tab:fitted-parameters}.}
\label{fig:posterior}
\end{figure*}

The joint posterior distribution of each $U_M$ and its $\muM$ (Fig.~\ref{fig:posterior}) shows a strong correlation between the two.
Raising $U_M$ changes the energy of each compound of $M$ by roughly the same amount per atom of $M$, and a change in $\muM$ cancels this common change.
The experimental values therefore constrain $U_M$ only through the part of its effect that differs from one compound of $M$ to another.

In the first fit (\ref{si-sn:all-metals}), the most probable value of $U_{\mathrm{Nb}}$ is zero, at the bound of the prior, and the posterior distribution of $U_{\mathrm{Cu}}$ has a single well whose tail keeps a large density at $U_{\mathrm{Cu}} = 0$.
The experimental values thus do not distinguish a positive $U$ from none for either metal.
The 95\% interval of $U_{\mathrm{Ag}}$ spans most of the prior range of 0 to 8~eV.
The enthalpies of the Re oxides deviate from the calculated values by similar amounts, which a free $\delta h_{\mathrm{Re}}$ would absorb and $U_{\mathrm{Re}}$ would not.
These four metals are therefore left out of the fit, with no $U_M$, no $\muM$ and none of the enthalpies of their compounds.
The four metals were chosen after the results of the first fit had been seen.
The enthalpies left out with them include those of ternary oxides that also contain Cr, Fe, Mo or W.
The values of the other twelve nevertheless change little, and the posterior median of each of their $U_M$ in the first fit lies within its 95\% interval in Table~\ref{tab:fitted-parameters}.

The posterior distributions of $U_{\mathrm{Co}}$, $U_{\mathrm{Ti}}$ and $U_{\mathrm{W}}$ have a second well at small $U$, and their 95\% intervals reach into it, down to \ULowCo{}, \ULowTi{} and \ULowW{}~eV.
For Co and Ti, the second well comes from a disagreement between the two kinds of experimental values, the enthalpies favouring a lower $U$ than the voltages.
The experimental values of Co make the Markov chain Monte Carlo sampling of $U_{\mathrm{Co}}$ unstable, and the lower end of its interval changes with the seed of the sampler.
For W, the second well has another origin.
The calculated formation enthalpies of most W oxides rise with $U_{\mathrm{W}}$ at nearly the same rate per W atom, and $\delta h_{\mathrm{W}}$ absorbs a shift common to all of them.
The value of $U_{\mathrm{W}}$ thus rests on small differences between a few W enthalpies, and the well at larger $U$ needs several of them together.
Leaving out the enthalpy of any one of WO$_2$, Na$_2$WO$_4$ and SrWO$_4$ removes the well at larger $U$, and leaving out that of BaWO$_4$ nearly removes the well at small $U$.

The intervals of Tc, Ru and Os are about 2~eV wide.
These metals rest on three, two and two enthalpies and no voltage (TcO$_2$, Tc$_2$O$_7$ and KTcO$_4$, RuO$_2$ and RuO$_4$, OsO$_2$ and OsO$_4$), and for Ru and Os, $U_M$ and $\muM$ reproduce both values exactly.
Their intervals therefore reflect $\tau_{\mathrm{H}}$ rather than any disagreement between the values.
The interval of $U_{\mathrm{Fe}}$, from \ULowFe{} to \UHighFe{}~eV, is the widest of the 3d metals, although it rests on many enthalpies and a voltage (Table~\ref{tab:fit-data}).
The Fe enthalpies change with $U_{\mathrm{Fe}}$ at nearly the same rate, and the posterior distribution is therefore flat over a wide range once $\delta h_{\mathrm{Fe}}$ absorbs the common shift.

The \FitVoltages{} cathode voltages are few against the \FitEnthalpies{} enthalpies, and they nevertheless determine $U$ for Co, Ti and Ni, whose posterior medians move when the voltages are left out of the fit (\ref{si-sn:no-voltage}).
Because a voltage is independent of $\muM$, it constrains $U_M$ directly, without the cancellation that limits what the enthalpies determine.
The enthalpies of LiNiO$_2$ and LaNiO$_3$ carry most of what constrains $U_{\mathrm{Ni}}$, and that of NiO almost none.
The voltage pulls $U_{\mathrm{Ni}}$ below the value the enthalpies favour, and the two are not independent, because LiNiO$_2$ is both an enthalpy in the fit and the lithiated end of the couple.
The enthalpies alone determine $U$ for V, Cr, Mn and Mo, whose binary oxides span several oxidation states, and their intervals are the narrowest in Table~\ref{tab:fitted-parameters}.
V and Mn also have cathode couples in the fit, and their posterior medians, like those of the other metals, change little when the voltages are left out.
Taken alone, the three Mn voltages favour a larger $U_{\mathrm{Mn}}$ than the enthalpies do.
Even at that value, the calculated voltage of one couple lies above experiment and those of the other two below it, each by more than its experimental uncertainty.

Swathilakshmi et al.~\cite{Swathilakshmi2023} also fitted $U$ for \rscanu{} oxides.
Their values lie within the 95\% intervals of this work (Table~\ref{tab:fitted-parameters}) for Ti (2.3~eV), Mn (1.8~eV), Fe (3.1~eV), Co (1.8~eV) and Ni (2.1~eV), and below them for V (1.0~eV) and Cr (0.45~eV, where they recommend \rscan{} without $U$).
Of the five inside, Ti, Fe, Co and Ni agree only weakly, because the intervals of this work for these metals are wide.
Mn agrees within an interval of this work that is among the narrowest of the 3d metals.
Their fit reads oxidation enthalpies between binary oxides of one metal rather than formation enthalpies of binary and ternary oxides, applies no O$_2$ correction, and reads no cathode voltages.
The difference for these two appears to be a systematic one that comes from the O$_2$ correction.
Without the O$_2$ correction, every computed oxidation enthalpy is higher by $\epsO$ per O$_2$ than with it.
The computed oxidation enthalpy rises with $U$, and a fit without the correction therefore meets the experimental enthalpy at a smaller $U$, which is the direction of the difference for V and Cr.

The posterior median of $U_{\mathrm{Fe}}$, \UFe{}~eV, is the largest in Table~\ref{tab:fitted-parameters}, and the value of Swathilakshmi et al.\ for Fe lies at the lower end of its interval.
The large value does not come from the cathode voltage, because the posterior median without the voltages is \UNoVoltageFe{}~eV (\ref{si-sn:no-voltage}).
Their smaller value is in the direction the O$_2$ correction gives, as for V and Cr.
The calculations differ as well, for example in the magnetic configurations, which they take from experiment and this work takes as the lowest in energy among those computed.

\subsection{Agreement with the fitted data}\label{sec:fit-agreement}

The correction scheme, at the fitted values of $U_M$ and $\muM$, lowers the mean absolute error (MAE) of the \ScoreEnthalpies{} enthalpies the fit reads from \ScoreMAEEnthalpyRscan{}~eV per atom with \rscan{} alone to \ScoreMAEEnthalpyScheme{}~eV per atom.
The corrected enthalpies are obtained from calculations of each compound at the fitted values of $U$, whereas the fit reads interpolated energies.
These are the values the parameters were fitted to, and the MAE measures how closely the fit reproduces them rather than how well it predicts other compounds.

\begin{figure*}[t]
\centering
\includegraphics[width=\textwidth]{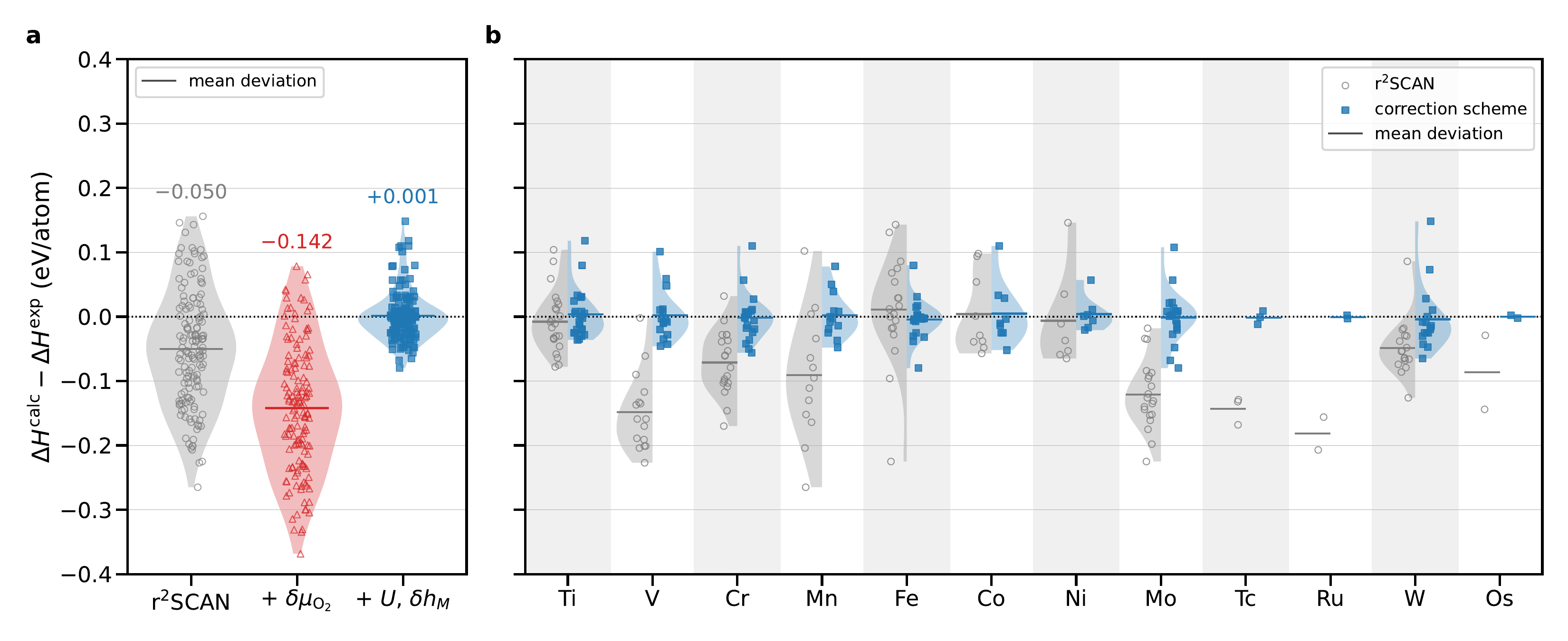}
\caption{\textbf{Agreement with the fitted enthalpies.}
\textbf{a}, Deviation of the calculated enthalpy from experiment for each formation enthalpy and enthalpy of reaction of Supplementary Tables~\ref{si-tab:fit-enthalpies} to~\ref{si-tab:fit-reactions}.
From left to right, each column adds one part of the corrections to the one before it, \rscan{} alone (open circles), \rscan{} with the O$_2$ correction $\epsO$ alone (open triangles), and the correction scheme at the fitted values of $U_M$ and $\muM$ (filled squares).
Each point is a compound, each violin the distribution of its column, and each bar its mean, whose value is printed above the violin.
\textbf{b}, The deviations of the compounds with a fitted metal, grouped by that metal, where a compound with two such metals appears under each.
For each metal, the left half-violin shows the deviations with \rscan{} alone and the right one those with the correction scheme, each with a bar at its mean and its compounds as points, and a half-violin is drawn only where the metal has enough compounds for a density.
The dotted lines mark agreement with experiment.}
\label{fig:fit-scores}
\end{figure*}

Figure~\ref{fig:fit-scores}a separates the two parts of the corrections, with the deviation of a compound taken as its calculated minus its experimental enthalpy and each compound counted once.
On the main-group oxides the O$_2$ correction is calibrated on, the deviation of \rscan{} is almost entirely the error of O$_2$, which the correction removes (\nameref{sec:o2-correction-results}).
On the enthalpies here, the same error makes \rscan{} less negative than experiment by \ScoreMeanOtwo{}~eV per atom on average, and the O$_2$ correction alone removes this part.
What it leaves is a mean deviation of $\ScoreMeanEps$~eV per atom, larger than the error of O$_2$ and opposite in sign.
The mean deviation of \rscan{} alone, $\ScoreMeanRscan$~eV per atom, is thus the sum of two errors of opposite sign that partly cancel, and its small size does not mean that \rscan{} describes these compounds well.
The values of $U_M$ and $\muM$ remove the error left by the O$_2$ correction, bringing the mean deviation to $\ScoreMeanScheme$~eV per atom.

Figure~\ref{fig:fit-scores}b groups the deviations by the metal with a fitted $U$ each compound contains.
With \rscan{} alone, the mean deviation differs from metal to metal.
With the correction scheme, the mean deviation of every metal is close to zero, as a fitted $\muM$ makes it.
The scheme thus removes the error of \rscan{} that differs between metals.

\begin{figure}[t]
\centering
\includegraphics[width=\linewidth]{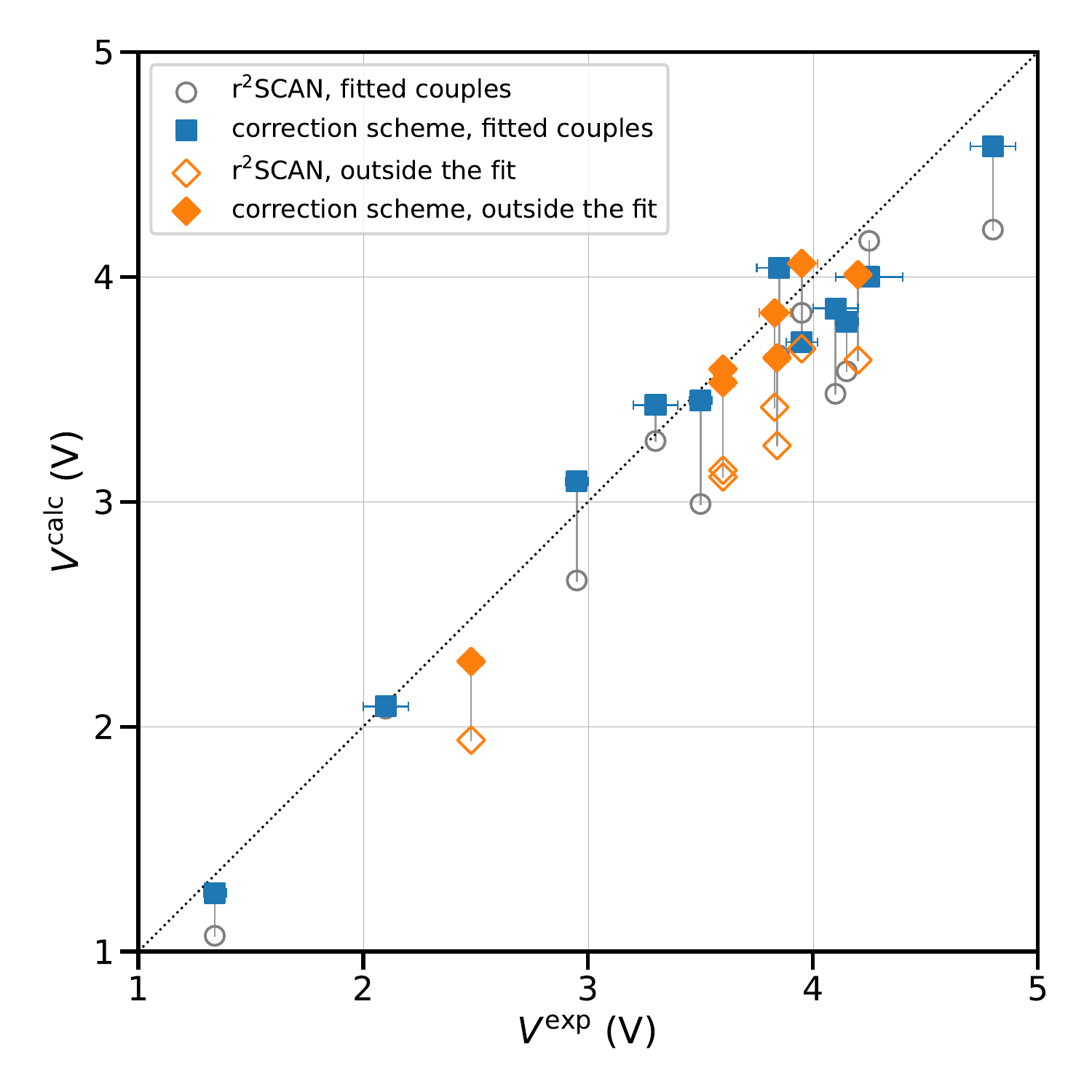}
\caption{\textbf{Cathode voltages.}
Calculated against experimental average voltage of each cathode couple the fit reads (blue squares, Supplementary Table~\ref{si-tab:fit-voltages}) and of each couple outside the fit (orange diamonds, Supplementary Table~\ref{si-tab:outside-voltages}).
Open markers are \rscan{} alone and filled markers the correction scheme at the fitted values of $U$, and the dotted line marks agreement with experiment.
A vertical line joins the two calculated voltages of each couple, and the horizontal bars are the experimental uncertainties.}
\label{fig:voltages}
\end{figure}

Because the corrections cancel in a voltage, only the values of $U$ change the calculated voltages.
Over the \ScoreVoltages{} cathode voltages the fit reads, the fitted values of $U$ lower the MAE from \ScoreMAEVoltageRscan{}~V with \rscan{} alone to \ScoreMAEVoltageScheme{}~V (Fig.~\ref{fig:voltages}).
Every \rscan{} voltage lies below experiment, with a mean deviation of $\ScoreMeanVoltageRscan$~V, and the fitted values of $U$ take the mean deviation to $\ScoreMeanVoltageScheme$~V and improve \ScoreVoltageImproved{} of the \ScoreVoltages{} couples.

\subsection{Cathode voltages outside the fit}\label{sec:cathode-voltages}

The \OutsideVoltages{} cathode couples of this subsection were collected from the electrochemistry literature after the values of $U$ were fixed, and no fitted parameter depends on them (\nameref{sec:methods-outside-couples}).
Four of these hosts contain sulfate or fluoride, which no compound of the fit contains, and the other three are phosphates in frameworks other than the olivine of the fitted phosphate couples.
Their transition metals are Fe, V and Ti, whose values of $U$ the fit determines, and the couples therefore test whether these values transfer to other compounds of the same metals, not to another metal.

With \rscan{} alone, every one of these voltages lies below experiment, with a mean deviation of $\OutsideMeanRscan$~V, as the voltages of the fitted couples do.
The fitted values of $U$ lower the MAE by a factor of \OutsideMAEFactor{}, from \OutsideMAERscan{}~V to \OutsideMAEScheme{}~V, on voltages that none of the fitted parameters depends on (Fig.~\ref{fig:voltages}).
They also take the mean deviation to $\OutsideMeanScheme$~V and improve every couple.
The values of $U$ fitted without the cathode voltages (\ref{si-sn:no-voltage}) give an MAE of \OutsideMAENoVoltage{}~V, with the voltages at the values of $U$ that differ from those of Table~\ref{tab:fitted-parameters} interpolated between the computed values of $U$, and the agreement thus does not depend on fitting voltages.

The calculated band gaps, to which no value of $U$ was fitted, were also compared with the experimental gaps compiled by Zhuo et al.~\cite{Zhuo2018} from earlier compilations~\cite{Kiselyova2016,Strehlow1973,Joshi1990,Madelung2004}, reduced to one value per composition in \code{matbench}~\cite{Dunn2020} and assigned Materials Project identifiers by Kingsbury et al.~\cite{Kingsbury2022}, and the values of $U$ do not degrade them (\ref{si-sn:band-gaps}).

\section{Discussion}\label{sec:discussion}

A fitted value of $U$ cannot be separated from the correction scheme and the fitting procedure that produced it, because adjusting $U$ in the fit absorbs whatever part of the error the other terms of the scheme leave.
The values of Swathilakshmi et al.~\cite{Swathilakshmi2023} for V and Cr, fitted without an O$_2$ correction, differ from those of this work in this way (\nameref{sec:fitted-parameters}).
The error of \rscan{} in the energy of the O$_2$ molecule enters every formation enthalpy and no cathode voltage, because the lithiated and delithiated compounds of a couple differ by lithium atoms alone.
Without an O$_2$ correction, a $U$ fitted to both would therefore be shifted to compensate for this error in the enthalpies, away from the value that the voltages favour.
A value of $U$ is therefore meaningful only together with the scheme and the procedure, and the soundness of the procedure has to be established on its own.
The correction scheme of this work builds on that of earlier GGA fits, which correct the overbinding of the O$_2$ molecule with an energy shift~\cite{Wang2006} and add a per-metal shift beside $U$~\cite{Jain2011}.
In this work, the procedure applied unchanged to PBE energies gives values close to those of these fits (\ref{si-sn:gga-u}), which supports its soundness.
The procedure does not depend on the functional, and it can be applied to another functional once the same compounds are computed with it.
Because the PBE fit reads the same data with the same procedure, the difference between its values of $U$ and those of \rscan{} comes from the functional.
A value of $U$ is defined for one functional, and only the direction of this difference is therefore compared.
The posterior median of $U_M$ is smaller in \rscan{} than in PBE for most metals, and no metal has a larger $U_M$ in \rscan{} with 95\% intervals that do not overlap.
This is the direction expected from the smaller self-interaction error of the meta-GGA, and agrees with the smaller $U$ that SCAN needs than the GGA~\cite{Long2020}.

The fitted values of $U$ are empirical corrections to energies, not measures of the on-site interaction of localized electrons.
They are therefore not comparable with a $U$ computed from first principles, which is the screened interaction within a chosen set of orbitals, and Table~\ref{tab:fitted-parameters} is compared only with other fitted values.
Nor does the fit constrain any property other than energies.
Of such properties, only the band gaps were checked, and the fitted values of $U$ do not degrade them (\ref{si-sn:band-gaps}).

The temperature dependence that the fit does not consider (\nameref{sec:data}) has been measured for related reactions.
For the enthalpies of reaction of ternary oxides from binary oxides, Hautier et al.\ found that integrating the heat capacity from 0 to 298~K changes the experimental values by 12~meV per atom on average, and that the zero-point energy is typically below a few meV per atom~\cite{Hautier2012}.
In a formation enthalpy from the unary phases, the contribution of the O$_2$ gas is the same for every molecule and is absorbed into $\epsO$, as its pressure--volume ($pV$) term is.
Contributions of the size Hautier et al.\ report are smaller than the posterior median of the model error $\tau_{\mathrm{H}}$, \TauEnthalpy{}~meV per atom.

Fitting one value of $U$ to all compounds of a metal is a crude approximation, because the value that suits a compound in one oxidation state need not suit a compound in another.
Accordingly, the posterior distributions of many metals are broad, and those of Co, Ti and W have a second well (\nameref{sec:fitted-parameters}).
Likewise, the experimental values the fit reads are all of bulk compounds, and the fitted value need not suit a metal atom at a surface, for example in an adsorption energy.
A $U$ that follows each compound would instead have to be computed for every structure of a high-throughput calculation, for example by linear response~\cite{Bastonero2025}, which is at present too costly.
Machine-learning models that predict linear-response values of $U$ from the structure~\cite{Uhrin2025,Xia2024} may make such a calculation affordable.

The scheme corrects oxides alone, and extending it to another anion needs enough experimental formation enthalpies of its compounds to determine $U$.
For fluorides, these are too few, because most metals have a measured binary fluoride in one oxidation state only, and the per-metal shift $\muM$ then absorbs the whole effect of $U$.
Likewise, too few experimental formation enthalpies of the oxides of the lanthanoids and actinoids are available at present for a fit to constrain a $U$ on their f shells.

\section{Methods}\label{sec:methods}

\subsection*{DFT calculations}\label{sec:methods-dft-calculations}

DFT calculations were performed with plane-wave basis sets and the projector augmented-wave (PAW) method~\cite{Blochl1994,Kresse1999} with the \rscan{} meta-generalized gradient approximation~\cite{Furness2020} and collinear spin polarization, as implemented in the Vienna ab initio simulation package (\code{VASP})~\cite{Kresse1993,Kresse1996a,Kresse1996b}.
The input parameters are those of the \rscan{} calculations of Shinagawa et al.~\cite{Shinagawa2026}, which are nearly identical to those of the MatPES dataset~\cite{Kaplan2025}, with the tags that set the Hubbard $U$ added, and are listed in Supplementary Table~\ref{si-tab:incar}.
The Hubbard correction is applied in the rotationally invariant form of Dudarev et al.~\cite{Dudarev1998} to the d states of each transition metal for which a $U$ is fitted, with no exchange parameter $J$.
The PBE\_64 PAW pseudopotential for each element is also chosen as by Shinagawa et al.\ and is listed in Supplementary Table~\ref{si-tab:paw-pseudopotentials}.
The fitted parameters apply to energies computed with these pseudopotentials.
A relaxation that did not converge with these parameters was repeated with SYMPREC $= 10^{-5}$, with POTIM $= 0.1$ or with IBRION $= 1$.
\subsection*{Unary phases}\label{sec:methods-unary-phases}

We compute each unary phase in the structure that the \emph{CRC Handbook of Chemistry and Physics} lists at 298.15~K and 1~atm~\cite{Rumble2025CRC}, starting from the observed Materials Project entry~\cite{Jain2013} in its space group that is lowest in energy above the convex hull, and relax it with the settings of \nameref{sec:methods-dft-calculations}.
The $\alpha$ and $\beta$ phases of boron share a space group, and the entry of the $\beta$ phase is taken.
Barium is relaxed from an exact body-centred cubic cell, because a relaxation from the Materials Project cell lost its cubic symmetry.
The structure and the Materials Project entry of each unary phase are listed in Supplementary Table~\ref{si-tab:unary-phases}.

The unary phases of Cr, Mn, Fe, Co and Ni are computed in several collinear magnetic configurations, and the configuration lowest in energy is taken.
Their initial moments are taken from tabulated atomic values with a lower bound of 0.6~$\mu_{\mathrm{B}}$, where $\mu_{\mathrm{B}}$ is the Bohr magneton, because a calculation started with zero moment remains nonmagnetic.
The other unary phases are computed in one configuration, started from 0.6~$\mu_{\mathrm{B}}$ on every atom.
Ti, V, Cu, Zn, Zr, Nb, Mo, Tc, Ru, Ag, W, Re and Os are not magnetically ordered, and their calculations end with zero moment.

Experimentally, Fe, Co and Ni are ferromagnetic, and Cr orders as an incommensurate spin-density wave~\cite{Blundell2001,PerezMato2012}.
We compute each of these four metals in the ferromagnetic configuration and in one antiferromagnetic configuration chosen for the calculation.
For body-centred cubic Cr and Fe, the antiferromagnetic configuration has the corner and body-centre atoms of the cubic cell antiparallel, which makes every atom antiparallel to its eight nearest neighbours.
For face-centred cubic Ni, it is the type-I configuration, in which ferromagnetic (001) planes alternate in sign, and for hexagonal close-packed Co, it has ferromagnetic (0001) planes alternating in sign.
For Cr, this configuration is the collinear approximation to the spin-density wave.
Fe, Co and Ni are lowest in the ferromagnetic configuration, in agreement with their well-known ferromagnetism.
Cr is lowest in the antiferromagnetic configuration.

The magnetic structure of $\alpha$-Mn is noncollinear~\cite{Lawson1994}, and no collinear configuration follows from it directly.
Its configurations are therefore searched.
The 29 atoms of its primitive cell fall into four symmetry-inequivalent sites, one on Wyckoff position $2a$, one on $8c$ and two on $24g$, and all eight configurations that assign a sign to the moments of each site, with the $2a$ moment positive, are computed.
The lowest of them is ferrimagnetic, with the moments of the $2a$ and $8c$ sites (3.6~$\mu_{\mathrm{B}}$) and of one $24g$ site (2.2~$\mu_{\mathrm{B}}$) parallel and those of the other $24g$ site (3.4~$\mu_{\mathrm{B}}$) antiparallel to them.

The O$_2$ molecule is computed in a 20~\AA{} box at the $\Gamma$ point, relaxing the atomic positions only.
The calculation starts from a magnetic moment of 1~$\mu_{\mathrm{B}}$ on each O atom, and the total moment is not constrained.

\subsection*{Formation enthalpies}\label{sec:methods-formation-enthalpies}

The formation enthalpy of an oxide is calculated with Eq.~\eqref{eq:uncorrected-formation-enthalpy}, from the total energies of the oxide, of its unary phases and of the O$_2$ molecule.
The pressure--volume ($pV$) term of a solid at atmospheric pressure is neglected, and a calculated formation enthalpy is a difference of total energies.
The $pV$ term of the O$_2$ gas, $k_{\mathrm{B}}T$ per molecule with the Boltzmann constant $k_{\mathrm{B}}$ and the temperature $T$, is the same for every O$_2$ molecule and is absorbed into the O$_2$ correction.
The temperature dependence of the enthalpies between 0 and 298.15~K, the zero-point energy included, is not considered (\nameref{sec:data}).
Experimental formation enthalpies, published in kJ per mole of the formula, are converted to eV per atom with 1~eV = 96.485~kJ\,mol$^{-1}$ and division by the number of atoms in the formula.

\subsection*{Enthalpies of reaction from binary oxides}\label{sec:methods-conversion-reactions}

Some experimental enthalpies are of reactions in which an oxide $P$ forms from binary oxides $A_i$ and O$_2$,
\begin{equation}
\sum_i a_i A_i + b\,\mathrm{O_2} \longrightarrow P ,
\end{equation}
where $a_i$ and $b$ are the stoichiometric coefficients of $A_i$ and O$_2$ for one formula of $P$, and $b$ is negative when O$_2$ is released.
The enthalpy of this reaction per atom of $P$ follows from the formation enthalpies of the oxides,
\begin{equation}
\Delta H_{\mathrm{c}} = \dHf(P) - \frac{1}{N_P} \sum_i a_i N_{A_i} \dHf(A_i) ,
\label{eq:conversion-enthalpy}
\end{equation}
where $N_P$ and $N_{A_i}$ are the numbers of atoms in the formulas of $P$ and $A_i$, and the formation enthalpy of O$_2$ is zero.
With the corrected formation enthalpies of Eq.~\eqref{eq:corrected-formation-enthalpy}, $\muM$ cancels for every metal and only $\epsO$ remains,
\begin{equation}
\Delta H_{\mathrm{c}}^{\mathrm{corr}}(U) = \Delta H_{\mathrm{c}}^{\mathrm{calc}}(U) - \frac{b}{N_P} \epsO ,
\end{equation}
where $\Delta H_{\mathrm{c}}^{\mathrm{calc}}(U)$ is Eq.~\eqref{eq:conversion-enthalpy} with the uncorrected values of Eq.~\eqref{eq:uncorrected-formation-enthalpy}.
Conversely, a measured $\Delta H_{\mathrm{c}}$ gives the formation enthalpy of $P$ through Eq.~\eqref{eq:conversion-enthalpy} with the experimental formation enthalpies of the binary oxides.

\subsection*{Lithium cathode voltages}\label{sec:methods-cathode-voltages}

A lithium cathode couple consists of a delithiated compound D and a lithiated compound L whose formulas differ by $n_{\mathrm{Li}}$ lithium atoms,
\begin{equation}
\mathrm{D} + n_{\mathrm{Li}}\,\mathrm{Li} \longrightarrow \mathrm{L} .
\end{equation}
Its average voltage follows from the formation enthalpies of the two compounds~\cite{Zhou2004},
\begin{equation}
V = \frac{N_{\mathrm{D}} \dHf(\mathrm{D}) - N_{\mathrm{L}} \dHf(\mathrm{L})}{n_{\mathrm{Li}} e} ,
\label{eq:voltage}
\end{equation}
where $N_{\mathrm{D}}$ and $N_{\mathrm{L}}$ are the numbers of atoms in the formulas of D and L, the formation enthalpy of lithium metal is zero, and $e$ is the elementary charge.
The calculated voltage $V^{\mathrm{calc}}(U)$ is Eq.~\eqref{eq:voltage} with the formation enthalpies of Eq.~\eqref{eq:uncorrected-formation-enthalpy}, in which the chemical potential of every element other than lithium cancels between the two compounds,
\begin{equation}
V^{\mathrm{calc}}(U) = \frac{E_{\mathrm{D}}(U) - E_{\mathrm{L}}(U) + n_{\mathrm{Li}} \mu_{\mathrm{Li}}}{n_{\mathrm{Li}} e} ,
\label{eq:voltage-energies}
\end{equation}
where $E_{\mathrm{D}}(U)$ and $E_{\mathrm{L}}(U)$ are the calculated total energies of the formulas of D and L, and $\mu_{\mathrm{Li}}$ is the calculated energy per atom of lithium metal.
The voltage of a couple that contains elements other than metals and O therefore requires the energy of no unary phase other than lithium metal.
The corrected voltage $V^{\mathrm{corr}}(U)$ is Eq.~\eqref{eq:voltage} with the formation enthalpies of Eq.~\eqref{eq:corrected-formation-enthalpy}.
The corrections $\muM$ and $\epsO$ cancel, because the two compounds contain the same amounts of oxygen and of every metal other than lithium, and lithium takes no shift,
\begin{equation}
V^{\mathrm{corr}}(U) = V^{\mathrm{calc}}(U) .
\end{equation}
A voltage therefore constrains $U_M$ alone, and the corrected voltage differs from that of \rscan{} alone, $V^{\mathrm{calc}}(0)$, through the Hubbard $U$ only.

\subsection*{Fitting the O\texorpdfstring{$_2$}{2} correction}\label{sec:o2-correction}

The O$_2$ correction is calibrated on \OtwoValues{} experimental formation enthalpies of the oxides selected in \nameref{sec:o2-correction-results}, \OtwoBinaryValues{} of binary oxides and \OtwoMulticationValues{} of oxides with two or more cations, read from critical compilations of thermochemical data~\cite{Robie1995,Yungman1999,Cox1989,Chase1998,Lamoreaux1984,Parker1971}.
The values are standard formation enthalpies of crystalline phases at 298.15~K, and four of them are taken from calorimetric studies~\cite{Huntelaar1992Ba,Huntelaar1992Sr}, because the compilations rest on older measurements for those compounds.
Where the compilations disagree, the value is chosen compound by compound.
For MgO, CaO and BeO the value recommended by CODATA~\cite{Cox1989} is taken, because CODATA evaluates its values together as one consistent set.
For BaO the NIST-JANAF value~\cite{Chase1998} is taken over the NBS value~\cite{Wagman1982}, because the NIST-JANAF value rests on a calorimetric measurement that the NBS value does not include.
The set includes three peroxides, in which oxygen is bonded to oxygen, because leaving them out changes the O$_2$ correction by less than 0.01~eV per O$_2$ molecule.
The values, their sources and the calculated formation enthalpies are listed in Supplementary Table~\ref{si-tab:o2-calibration}.

Each compound is computed in the structure of an observed Materials Project entry~\cite{Jain2013} in the space group of the polymorph whose enthalpy was measured, choosing the entry lowest in energy above the convex hull among them, and polymorphs with their own experimental values are computed separately.
The structures are relaxed with the settings of \nameref{sec:methods-dft-calculations} without a Hubbard $U$, and, having no magnetic ions, each in one configuration started from 0.6~$\mu_{\mathrm{B}}$ on every atom.
The formation enthalpies are calculated with Eq.~\eqref{eq:uncorrected-formation-enthalpy}, with the \rscan{} energy of each oxide.

Because no metal in these oxides takes a Hubbard $U$, Eq.~\eqref{eq:corrected-formation-enthalpy} reduces for them to $\dHf^{\mathrm{corr}} = \dHf^{\mathrm{calc}} - x_{\mathrm{O_2}} \epsO$, where $x_{\mathrm{O_2}} = n_{\mathrm{O}}/2N$ is the number of O$_2$ molecules in the formation reaction per atom of the oxide.
The correction is estimated by the mean, over the $K$ calibration values, of the shift that brings each calculated value to experiment,
\begin{equation}
\epsO \approx \frac{1}{K} \sum_{i=1}^{K} \frac{\dHf^{\mathrm{calc}}(i) - \dHf^{\mathrm{exp}}(i)}{x_{\mathrm{O_2}}(i)} ,
\label{eq:o2-correction}
\end{equation}
where $i$ labels the calibration values and $\dHf^{\mathrm{exp}}(i)$ is the experimental formation enthalpy.
The experimental uncertainties do not enter.
The uncertainty of $\epsO$ is the standard deviation of the differences averaged in Eq.~\eqref{eq:o2-correction}.

\subsection*{Experimental data preparation for transition-metal oxides}\label{sec:methods-experimental-data}

The experimental formation enthalpies are standard formation enthalpies of crystalline oxides at 298.15~K, taken from the compilation of Wang et al.~\cite{Wang2021} and from the NBS tables~\cite{Wagman1982}, and where both carry a compound the value of Wang et al.\ is taken.
The values of the NBS tables are read from the dataset that the National Institute of Standards and Technology publishes of them~\cite{Reed2020}.
The compilation of Wang et al.\ takes each value from an earlier source~\cite{Kubaschewski1993,Chase1998,LandoltBornstein,Grindy2013}, which Supplementary Table~\ref{si-tab:fit-enthalpies} cites beside the compilation.
We check these values against additional sources, namely critical evaluations and measurements of thermochemical data~\cite{Chase1998,Yungman1999,Robie1995,Cox1989,Lamoreaux1984,Wagman1969,Wagman1981,Lemire2013,Rard1999,ONeill1997,Chatillon2018,Arjmand2015}.
Where the additional sources agree on a value other than that of the compilation, their value is taken.
Compounds that neither compilation carries are added from the additional sources.
A compound is left out when its existence as a crystalline phase at 298.15~K is not confirmed, as for Ni$_2$O$_3$, KVO$_4$ and Na$_2$MnO$_4$, or when the measured phase is not the stoichiometric compound that is computed, as for TiO$_{1\pm\delta}$.
A compound is also left out when its values disagree with each other, or with those of related compounds, beyond what the evaluations reconcile, as for Ba$_3$WO$_6$ against Ca$_3$WO$_6$.
Each value carries the uncertainty its source states, and the NBS tables state none.
The formation enthalpies the fit reads as measured are listed with their sources, the structures they are computed in and the calculated values in Supplementary Table~\ref{si-tab:fit-enthalpies}.

The enthalpies of LiCoO$_2$, LiNiO$_2$ and LiMnO$_2$ are measured by drop-solution calorimetry as enthalpies of reaction from the binary oxides (\nameref{sec:methods-conversion-reactions})~\cite{Wang2004LiMO2,Wang2005LiMnO2}.
Lithium is a main-group element, and Li$_2$O is among the oxides the O$_2$ correction is calibrated on (\nameref{sec:o2-correction}).
These three enthalpies are therefore converted to formation enthalpies with Eq.~\eqref{eq:conversion-enthalpy} and the experimental values of the binary oxides, and are listed with both values in Supplementary Table~\ref{si-tab:fit-conversions}.
The enthalpies of LaCoO$_3$ and LaNiO$_3$ are measured in the same way~\cite{Cheng2005}.
Lanthanum carries no correction of its own, and the fit reads these two as their enthalpies of reaction from the binary oxides, in which the calculated energy of La$_2$O$_3$ enters and the error of lanthanum largely cancels.
They are listed in Supplementary Table~\ref{si-tab:fit-reactions}.

The cathode voltages are average open-circuit voltages over the lithium range of each couple (\nameref{sec:methods-cathode-voltages}), read from the electrochemistry literature~\cite{Zhou2004,Whittingham1976,Reimers1992,Delmas1994,Colbow1989}, and the fit reads eleven couples, of layered, spinel and olivine hosts and of the sulfide TiS$_2$.
The couples are listed with their sources and the calculated voltages in Supplementary Table~\ref{si-tab:fit-voltages}.

\subsection*{Computed data preparation for transition-metal oxides}\label{sec:methods-computed-data}

We compute each compound in the crystal structure of a Materials Project entry~\cite{Jain2013} of the phase whose enthalpy was measured.
For a compound taken from the compilation of Wang et al., we use the entry assigned there~\cite{Wang2021}.
Wang et al. chose it among the non-theoretical entries within 200~meV per atom of the convex hull with a matching structure in the Inorganic Crystal Structure Database (ICSD)~\cite{Zagorac2019}, as the one lowest above the hull in the space group the source reports.
For the other compounds, we take the entry marked as experimentally observed, preferring one with an ICSD record when there are several, and a compound with no observed entry is not computed.
Where the space group of the measured phase is reported, we compare the entry with it and take an observed entry in that space group when the two differ.

The lithiated and delithiated compounds of each cathode couple (\nameref{sec:methods-cathode-voltages}) are computed at the same value of $U$, each in the structure of an observed Materials Project entry of that compound.
MnPO$_4$ has no observed entry and is computed in the olivine framework of LiMnPO$_4$ with the lithium removed.
The fully delithiated CoO$_2$ and NiO$_2$ each have two observed structures, which stack the same MO$_2$ layers differently, the O1 stacking in space group $P\bar{3}m1$ and the O3 stacking of the lithiated compound in $R\bar{3}m$, in the notation of Delmas et al.~\cite{Delmas1980}.
Each is computed in both structures, and the lower of the two energies is taken at each value of $U$.

The compounds containing magnetic ions, which are metal ions with a partly filled d shell at their formal oxidation state, are computed in several collinear magnetic configurations, and the configuration lowest in energy is taken.
The other compounds are computed in one configuration, started ferromagnetic.
Initial moments are set as for the unary phases (\nameref{sec:methods-unary-phases}).
For a binary oxide with a magnetic ion, one more configuration is computed, started from the magnetic moments of its Materials Project entry, and is among those of which the lowest is taken.

The configurations of a compound are enumerated on the sublattice of its magnetic ions with the \code{MagneticStructureEnumerator} class of \code{pymatgen}~\cite{Ong2013,George2026,Horton2019}, which builds on the derivative-structure enumeration of \code{enumlib}~\cite{Hart2008,Hart2012}.
Its default ferromagnetic, antiferromagnetic and ferrimagnetic strategies are used, and the supercell is up to twice the primitive cell when the primitive cell holds at most two magnetic ions.
Of the enumerated configurations, symmetry-equivalent ones are removed, and at most six are kept per compound in the order the enumeration returns them, with the ferromagnetic configuration always kept.
A compound with two kinds of magnetic ions is enumerated on both sublattices together, and W and Mo are taken as hexavalent, and therefore nonmagnetic, in a compound with another transition metal.
When the primitive cell holds more than eight magnetic ions, the enumeration is not run, because the number of configurations grows rapidly with the number of magnetic ions.
Such a compound is computed in the ferromagnetic configuration and in up to two simple configurations with alternating signs, one alternating between the symmetry-inequivalent sites of the magnetic ion and one between successive atomic layers along the c axis.
Their purpose is to ensure that each such compound is also computed in a configuration that is not ferromagnetic, rather than to reproduce its experimental magnetic order.

The total energy $E$ of a compound in Eq.~\eqref{eq:uncorrected-formation-enthalpy} depends on the Hubbard $U_M$ of its metals.
We write it $E(U_M)$ for a compound with one metal $M$ for which a $U$ is fitted, and $E(U_M, U_{M'})$ for a compound with two, such as MnFe$_2$O$_4$.
It is computed with the Hubbard correction of \nameref{sec:methods-dft-calculations}, and the structure is relaxed separately at each value of $U$.
The energy $E(U_M)$ is computed at $U_M = 0$, 1, 2 and 4~eV, and also at 6~eV for Fe, W and Ag.
The energy $E(U_M, U_{M'})$ is computed with $U_M$ and $U_{M'}$ varied independently, at every combination of their values, each taking 0, 2, 4 and 6~eV for Fe and Ag and 0, 1, 2 and 4~eV for the other metals.
Four compounds the fit reads lack some of these values of $U_M$, because their calculations there did not converge, SrTiO$_3$ being computed only at 1 and 4~eV, Na$_4$V$_2$O$_7$ without 2~eV, and K$_2$Cr$_3$O$_{10}$ and K$_2$Mo$_4$O$_{13}$ without 4~eV.

For $E(U_M)$, every magnetic configuration is computed at every value of $U_M$.
For $E(U_M, U_{M'})$, fewer magnetic configurations are computed, to keep the number of calculations tractable.
Every configuration is computed only with \rscan{} alone, at $U_M = U_{M'} = 0$.
At the remaining combinations, only two configurations are computed, the one lowest in energy with \rscan{} alone and, if that one is ferromagnetic, the lowest non-ferromagnetic one, or otherwise the ferromagnetic one.
The four wolframites $M$WO$_4$ with $M$ = Mn, Fe, Co and Ni are an exception to this rule.
In preliminary calculations of them with one value of $U$ for both metals, configurations of the same kind lay within a few meV per $M$WO$_4$ of each other, and which of them was lowest changed with $U$.
For these four compounds, a third configuration is therefore also computed, the next lowest of the same kind as the lowest one, which lies within 30~meV per $M$WO$_4$ of it with \rscan{} alone.

\subsection*{Energies as functions of \texorpdfstring{$U$}{U}}\label{sec:methods-energy-curves}

The fit of $U_M$ and $\muM$ varies $U_M$ continuously, whereas the energies are computed at a few values of $U_M$ only (\nameref{sec:methods-computed-data}).
At each computed value of $U_M$, or combination of $U_M$ and $U_{M'}$, the energy is taken as the lowest over the magnetic configurations computed there.
The configuration lowest in energy can differ from one value of $U$ to another, and the resulting function has a kink where one configuration takes over from another.

Other studies that fitted $U$ to experimental enthalpies have computed the energy at intervals of 1~eV or less in $U$ and joined consecutive values by straight lines~\cite{Wang2006,SaiGautam2018,Long2020,Swathilakshmi2023}, or have fitted one linear function of $U$ through them~\cite{Artrith2022}.

Between computed values, the function is interpolated through these energies rather than fitted to them, because a fitted function would round off the kink and shift the energies at the computed values of $U$.
The function $E(U_M)$ is interpolated by a piecewise cubic polynomial with a continuous first derivative~\cite{Fritsch1984}, as implemented in \code{SciPy}~\cite{Virtanen2020}, rather than by straight segments.
It rises or falls wherever the computed energies rise or fall and stays between the energies at the ends of each interval, and therefore does not overshoot at a kink.

For a ternary oxide with two metals for which a $U$ is fitted, $E(U_M, U_{M'})$ is interpolated bilinearly.
Between two consecutive computed values of $U_M$ and two of $U_{M'}$, it is $a + b\,U_M + c\,U_{M'} + d\,U_M U_{M'}$, with the four coefficients set by the energies computed at the four pairs of these values.
The reaction energy of each cathode couple is interpolated by straight segments.
Interpolating all three by the piecewise cubic polynomial changes no posterior median of $U$ by more than 0.06~eV.
Above the largest computed $U$, each function is extended linearly in each $U$, which lets the fit reach values of $U$ beyond those computed.

For SrTiO$_3$, which has two computed values, $E(U_M)$ is a straight line.
For Fe$_2$Mo$_3$O$_{12}$, which lacks two of the combinations, $E(U_M, U_{M'})$ is fitted by a sum of quadratic functions of $U_M$ and of $U_{M'}$ rather than interpolated.

\subsection*{Fitting \texorpdfstring{$U_M$ and $\muM$}{U and delta h}}\label{sec:fitting-u-mu}

The model errors $\tau_{\mathrm{H}}$ and $\tau_V$ are inferred with the other parameters, as Tavadze et al.\ infer the variance of each property in their calibration of $U$~\cite{Tavadze2021}.
Where the source states no uncertainty, which is the case for 57 of the \FitEnthalpies{} enthalpies and for none of the voltages, $s_{\mathrm{f}}(x)$ or $s_{\mathrm{c}}(x)$ is set to the mean of the stated uncertainties per atom, 9.5~meV per atom.

For the priors of Eq.~\eqref{eq:posterior}, we take each $U_M$ to be uniform on 0 to 8~eV in $p(U)$, each $\muM$ to be uniform over all real values in $p(\delta h)$, and $\tau_{\mathrm{H}}$ and $\tau_V$ to be half-normal with scales of 0.1~eV per atom and 0.3~V in $p(\tau)$.

Before sampling, $\delta h$ is integrated out of the posterior of Eq.~\eqref{eq:posterior} analytically, which leaves the posterior $p(U, \tau \mid \mathcal{X}, \mathcal{Y}) \propto p(U)\, p(\tau)\, p(\mathcal{Y} \mid \mathcal{X}, U, \tau)$ of $U$ and $\tau$.
This is possible because the corrected calculated values are linear in $\delta h$ and $p(\delta h)$ is uniform.
Let $x_i$, $y_i$, $f_i$ and $\sigma_i^2$ be the structures, the experimental value, the corrected calculated value and the variance that Eq.~\eqref{eq:observation-model} assigns to the $i$th experimental value, for example $\dHf^{\mathrm{exp}}(x_i)$, $\dHf^{\mathrm{corr}}(x_i, U, \delta h)$ and $s_{\mathrm{f}}(x_i)^2 + \tau_{\mathrm{H}}^2$ for a formation enthalpy, with $\mathcal{X} = \{x_i\}$ and $\mathcal{Y} = \{y_i\}$.
Then $f_i = g_i(U) + \sum_M a_{iM} \muM$, where $g_i(U)$ is the corrected calculated value with every $\muM$ set to zero, a function of $U$ through the interpolated energies (\nameref{sec:methods-energy-curves}), and $a_{iM} = n_M/N$ for a formation enthalpy and $a_{iM} = 0$ for an enthalpy of reaction compared directly and for a voltage.
The integral of the likelihood $p(\mathcal{Y} \mid \mathcal{X}, U, \delta h, \tau)$ over $\delta h$ is
\begin{equation}
\begin{split}
\ln p(\mathcal{Y} \mid \mathcal{X}, U, \tau) \\
= {} & -\frac{1}{2} \bigl\lVert (\bm{I} - \bm{P})\, \bm{d} \bigr\rVert^2 - \sum_i \ln \sigma_i \\
& - \frac{1}{2} \ln \det\bigl(\bm{B}^{\top} \bm{B}\bigr) \\
& - \frac{n_y - n_{\delta h}}{2} \ln 2\pi ,
\end{split}
\label{eq:marginal-likelihood}
\end{equation}
where the vector $\bm{d}$ and the matrix $\bm{B}$ have the components $d_i = (g_i(U) - y_i)/\sigma_i$ and $B_{iM} = a_{iM}/\sigma_i$, $\bm{I}$ is the identity matrix, $\bm{P} = \bm{B} (\bm{B}^{\top} \bm{B})^{-1} \bm{B}^{\top}$ projects onto the columns of $\bm{B}$, and $n_y$ and $n_{\delta h}$ are the numbers of experimental values and of per-metal shifts.
Given $U$ and $\tau$, the posterior of $\delta h$ is
\begin{equation}
\begin{split}
& \delta h \mid \mathcal{X}, \mathcal{Y}, U, \tau \\
& \quad \sim \mathcal{N}\bigl(-(\bm{B}^{\top} \bm{B})^{-1} \bm{B}^{\top} \bm{d},\ (\bm{B}^{\top} \bm{B})^{-1}\bigr) ,
\end{split}
\label{eq:shift-posterior}
\end{equation}
whose mean is the weighted least-squares solution for $\delta h$.
The reported values of $\muM$ are this mean at the rounded $U_M$, with $\tau$ at the maximum of the posterior.

The posterior $p(U, \tau \mid \mathcal{X}, \mathcal{Y})$ is sampled with the likelihood of Eq.~\eqref{eq:marginal-likelihood}.
Before sampling, the posterior is maximized with Powell's method as implemented in \code{SciPy}~\cite{Virtanen2020} from 128 starting points drawn uniformly over the prior, and every start reaches the same maximum.
Eight chains, started near this maximum, are sampled with the differential-evolution Markov chain sampler \code{DEMetropolisZ}~\cite{terBraak2008} of \code{PyMC}~\cite{AbrilPla2023}, which requires no gradient of the interpolated energies, each with 30,000 draws after 30,000 tuning steps.

Draws from the posterior $p(U, \delta h, \tau \mid \mathcal{X}, \mathcal{Y})$ of Eq.~\eqref{eq:posterior} are obtained without sampling $\delta h$.
At every 20th draw of $U$ and $\tau$, one vector $\delta h$ is drawn from its conditional posterior of Eq.~\eqref{eq:shift-posterior}, which gives 12,000 draws of $U$, $\delta h$ and $\tau$ together.
Because Eq.~\eqref{eq:shift-posterior} gives the conditional posterior exactly, each of these draws follows the joint posterior, and no additional Markov chain Monte Carlo sampling of $\delta h$ is needed.

The convergence of the chains is checked with two diagnostics of Vehtari et al.~\cite{Vehtari2021}, each computed separately from the draws of every $U_M$.
The rank-normalized $\hat{R}$ compares the spread of the draws of one $U_M$ between the chains with that within each chain, and approaches 1 when all the chains sample the same distribution.
Vehtari et al.\ recommend values below 1.01, and $\hat{R}$ is at most 1.005 for every value of $U$ except $U_{\mathrm{Co}}$, for which it is 1.028.
The bulk effective sample size, the number of independent draws that the correlated draws of the chains are worth, is at least 2,810 for every value of $U$ except $U_{\mathrm{Co}}$, for which it is 366.
The posterior of $U_{\mathrm{Co}}$ has a lower tail that few of the chains reach.

The standard deviation of the posterior of each $U_M$ is compared with that of its uniform prior, $8/\sqrt{12} \approx 2.31$~eV.
A value of $U$ is called weakly identified when its posterior standard deviation lies between half and three quarters of the prior one, 1.15 and 1.73~eV, and not identified when it exceeds three quarters.

\subsection*{Cathode couples outside the fit}\label{sec:methods-outside-couples}

The cathode voltages scored under \nameref{sec:cathode-voltages} are those of seven couples collected after the values of $U$ were fixed, read as the voltages the fit reads are (\nameref{sec:methods-experimental-data}) from the electrochemistry literature~\cite{Manthiram1989,Kerr2000,Reynaud2012,Delmas1988,Yin2003,Recham2010,Barker2003}.
They are computed as the couples the fit reads are (\nameref{sec:methods-computed-data}), except that two compounds start from entries of the Crystallography Open Database~\cite{Grazulis2012} and the lithiated Li$_2$Fe$_2$(SO$_4$)$_3$, whose structure has not been solved, is computed in a theoretical structure.
Their voltages at the fitted values of $U$ are obtained from calculations of both compounds at these values.
The couples are listed with their sources and the calculated voltages in Supplementary Table~\ref{si-tab:outside-voltages}.

\subsection*{Statistics and reproducibility}\label{sec:statistics}

Each value of $U$ is reported as the median of its posterior distribution (\nameref{sec:fitting-u-mu}) with the equal-tailed 95\% interval.
The uncertainty of $\epsO$ is the standard deviation of the per-O$_2$ differences it is the mean of (\nameref{sec:o2-correction}).
The formation enthalpies scored in Results are of compounds the fit also reads, and their errors are in-sample.
The seven cathode voltages scored in Results were collected after the values of $U$ were fixed, and no fitted parameter depends on them.
Every dataset and table is generated by scripts from the calculations and experimental values.
The Bayesian fit will be released as a package (\nameref{sec:code-availability}), which reads the Supplementary Data and evaluates the same likelihood as the scripts used in this work.
Because Markov chain Monte Carlo sampling is random in nature, a repeated run is expected to reproduce the reported values within their sampling uncertainty, and an exact reproduction is not guaranteed.

\subsection*{Code availability}\label{sec:code-availability}

The code that fits $\epsO$, $U_M$ and $\muM$ to the Supplementary Data, a Python package, and the correction scheme with the fitted parameters, a \code{Compatibility} class of \code{pymatgen}~\cite{Ong2013}, will be released with the published version of this article.

\bmhead{Data availability}

The Supplementary Information and the Supplementary Data referred to in this article are not part of this preprint.
The experimental formation enthalpies compiled by Wang et al.~\cite{Wang2021} and the band gaps compiled by Zhuo et al.~\cite{Zhuo2018}, with the Materials Project identifiers assigned by Kingsbury et al.~\cite{Kingsbury2022}, are available through \code{matminer}~\cite{Ward2018}.
The experimental formation enthalpies of the NBS tables~\cite{Wagman1982} are available as a dataset of the National Institute of Standards and Technology~\cite{Reed2020}, which is the source of the values used here.
In this work, in 2026, these values were converted from kJ~mol$^{-1}$ to eV per atom, and the value of Cu$_2$Al$_2$O$_4$ was halved to give that of AlCuO$_2$.

\bmhead{Acknowledgements}

The authors thank Daiki Shintani, Shigeru Iwase, Iori Kurata and Yu Miyazaki for fruitful discussions and for their careful reading of the manuscript.

This study was conducted as part of preparations for the ``GENIAC (Generative AI Accelerator Challenge)'' project to be implemented by Japan's Ministry of Economy, Trade and Industry (METI) and the New Energy and Industrial Technology Development Organization (NEDO).

\bmhead{Author contributions}

C.S.\ contributed to conceptualization, methodology, software, formal analysis, investigation and writing (review and editing).
K.S.\ contributed to conceptualization, methodology, data curation, validation, visualization, writing (original draft) and writing (review and editing).
J.O.\ contributed to data curation, validation and writing (review and editing).
T.S.\ contributed to writing (review and editing).
S.T.\ contributed to resources, supervision and writing (review and editing).
J.L.\ contributed to supervision and writing (review and editing).

\bmhead{Competing interests}

C.S., K.S., J.O., T.S.\ and S.T.\ are employees of Preferred Networks, Inc., and J.L.\ is an advisor to Preferred Networks, Inc.
Matlantis Corp., a joint venture among Preferred Networks, Inc., ENEOS Corporation and Mitsubishi Corporation, offers Matlantis, a software-as-a-service platform, and plans to incorporate a product based on this research into it.

\bmhead{Use of AI-assisted technology}

The authors used the AI assistant Claude (Anthropic, Opus 5.5 and Fable 5) throughout this work, to write and run analysis code, to collate and check the experimental data against their sources, and to draft and revise the manuscript.
The authors reviewed its output and take full responsibility for the content of this manuscript.


\begin{thebibliography}{100}
\expandafter\ifx\csname url\endcsname\relax
  \def\url#1{\burl{#1}}\fi
\expandafter\ifx\csname urlprefix\endcsname\relax\def\urlprefix{URL }\fi
\providecommand{\bibinfo}[2]{#2}
\providecommand{\eprint}[2][]{\url{#2}}
\providecommand{\doi}[1]{\url{https://doi.org/#1}}
\bibcommenthead

\bibitem{Jain2013}
\bibinfo{author}{Jain, A.} \emph{et~al.}
\newblock \bibinfo{title}{Commentary: The {Materials Project}: A materials
  genome approach to accelerating materials innovation}.
\newblock \emph{\bibinfo{journal}{APL Mater.}} \textbf{\bibinfo{volume}{1}},
  \bibinfo{pages}{011002} (\bibinfo{year}{2013}).

\bibitem{Jain2016}
\bibinfo{author}{Jain, A.}, \bibinfo{author}{Shin, Y.} \&
  \bibinfo{author}{Persson, K.~A.}
\newblock \bibinfo{title}{Computational predictions of energy materials using
  density functional theory}.
\newblock \emph{\bibinfo{journal}{Nat. Rev. Mater.}}
  \textbf{\bibinfo{volume}{1}}, \bibinfo{pages}{15004} (\bibinfo{year}{2016}).

\bibitem{Zhou2004}
\bibinfo{author}{Zhou, F.}, \bibinfo{author}{Cococcioni, M.},
  \bibinfo{author}{Marianetti, C.~A.}, \bibinfo{author}{Morgan, D.} \&
  \bibinfo{author}{Ceder, G.}
\newblock \bibinfo{title}{First-principles prediction of redox potentials in
  transition-metal compounds with {LDA}+{U}}.
\newblock \emph{\bibinfo{journal}{Phys. Rev. B}} \textbf{\bibinfo{volume}{70}},
  \bibinfo{pages}{235121} (\bibinfo{year}{2004}).

\bibitem{Hohenberg1964}
\bibinfo{author}{Hohenberg, P.} \& \bibinfo{author}{Kohn, W.}
\newblock \bibinfo{title}{Inhomogeneous electron gas}.
\newblock \emph{\bibinfo{journal}{Phys. Rev.}} \textbf{\bibinfo{volume}{136}},
  \bibinfo{pages}{B864--B871} (\bibinfo{year}{1964}).

\bibitem{Kohn1965}
\bibinfo{author}{Kohn, W.} \& \bibinfo{author}{Sham, L.~J.}
\newblock \bibinfo{title}{Self-consistent equations including exchange and
  correlation effects}.
\newblock \emph{\bibinfo{journal}{Phys. Rev.}} \textbf{\bibinfo{volume}{140}},
  \bibinfo{pages}{A1133--A1138} (\bibinfo{year}{1965}).

\bibitem{Hafner2008}
\bibinfo{author}{Hafner, J.}
\newblock \bibinfo{title}{Ab-initio simulations of materials using {VASP}:
  Density-functional theory and beyond}.
\newblock \emph{\bibinfo{journal}{J. Comput. Chem.}}
  \textbf{\bibinfo{volume}{29}}, \bibinfo{pages}{2044--2078}
  (\bibinfo{year}{2008}).

\bibitem{Perdew1996}
\bibinfo{author}{Perdew, J.~P.}, \bibinfo{author}{Burke, K.} \&
  \bibinfo{author}{Ernzerhof, M.}
\newblock \bibinfo{title}{Generalized gradient approximation made simple}.
\newblock \emph{\bibinfo{journal}{Phys. Rev. Lett.}}
  \textbf{\bibinfo{volume}{77}}, \bibinfo{pages}{3865--3868}
  (\bibinfo{year}{1996}).

\bibitem{Perdew2001}
\bibinfo{author}{Perdew, J.~P.} \& \bibinfo{author}{Schmidt, K.}
\newblock \bibinfo{title}{Jacob's ladder of density functional approximations
  for the exchange-correlation energy}.
\newblock \emph{\bibinfo{journal}{AIP Conf. Proc.}}
  \textbf{\bibinfo{volume}{577}}, \bibinfo{pages}{1--20}
  (\bibinfo{year}{2001}).

\bibitem{Sun2015}
\bibinfo{author}{Sun, J.}, \bibinfo{author}{Ruzsinszky, A.} \&
  \bibinfo{author}{Perdew, J.~P.}
\newblock \bibinfo{title}{Strongly constrained and appropriately normed
  semilocal density functional}.
\newblock \emph{\bibinfo{journal}{Phys. Rev. Lett.}}
  \textbf{\bibinfo{volume}{115}}, \bibinfo{pages}{036402}
  (\bibinfo{year}{2015}).

\bibitem{Zhang2018}
\bibinfo{author}{Zhang, Y.} \emph{et~al.}
\newblock \bibinfo{title}{Efficient first-principles prediction of solid
  stability: Towards chemical accuracy}.
\newblock \emph{\bibinfo{journal}{npj Comput. Mater.}}
  \textbf{\bibinfo{volume}{4}}, \bibinfo{pages}{9} (\bibinfo{year}{2018}).

\bibitem{Furness2020}
\bibinfo{author}{Furness, J.~W.}, \bibinfo{author}{Kaplan, A.~D.},
  \bibinfo{author}{Ning, J.}, \bibinfo{author}{Perdew, J.~P.} \&
  \bibinfo{author}{Sun, J.}
\newblock \bibinfo{title}{Accurate and numerically efficient {r$^2$SCAN}
  meta-generalized gradient approximation}.
\newblock \emph{\bibinfo{journal}{J. Phys. Chem. Lett.}}
  \textbf{\bibinfo{volume}{11}}, \bibinfo{pages}{8208--8215}
  (\bibinfo{year}{2020}).

\bibitem{Kingsbury2022}
\bibinfo{author}{Kingsbury, R.} \emph{et~al.}
\newblock \bibinfo{title}{Performance comparison of {r$^2$SCAN} and {SCAN}
  {metaGGA} density functionals for solid materials via an automated,
  high-throughput computational workflow}.
\newblock \emph{\bibinfo{journal}{Phys. Rev. Mater.}}
  \textbf{\bibinfo{volume}{6}}, \bibinfo{pages}{013801} (\bibinfo{year}{2022}).

\bibitem{Kaplan2025}
\bibinfo{author}{Kaplan, A.~D.} \emph{et~al.}
\newblock \bibinfo{title}{A foundational potential energy surface dataset for
  materials} (\bibinfo{year}{2025}).
\newblock
  \bibinfo{eprint}{{\href{https://arxiv.org/abs/2503.04070}{{arXiv:2503.04070}}}}.

\bibitem{Kuner2025}
\bibinfo{author}{Kuner, M.~C.}, \bibinfo{author}{Kaplan, A.~D.},
  \bibinfo{author}{Persson, K.~A.}, \bibinfo{author}{Asta, M.} \&
  \bibinfo{author}{Chrzan, D.~C.}
\newblock \bibinfo{title}{{MP-ALOE}: an {r$^2$SCAN} dataset for universal
  machine learning interatomic potentials}.
\newblock \emph{\bibinfo{journal}{npj Comput. Mater.}}
  \textbf{\bibinfo{volume}{11}}, \bibinfo{pages}{352} (\bibinfo{year}{2025}).

\bibitem{Shinagawa2026}
\bibinfo{author}{Shinagawa, C.} \emph{et~al.}
\newblock \bibinfo{title}{{Matlantis-PFP} v8: Universal machine learning
  interatomic potential with better experimental agreements via {r$^2$SCAN}
  functional} (\bibinfo{year}{2026}).
\newblock
  \bibinfo{eprint}{{\href{https://arxiv.org/abs/2603.11063}{{arXiv:2603.11063}}}}.

\bibitem{SaiGautam2018}
\bibinfo{author}{Sai~Gautam, G.} \& \bibinfo{author}{Carter, E.~A.}
\newblock \bibinfo{title}{Evaluating transition-metal oxides within {DFT-SCAN}
  and {SCAN}+{U} frameworks for solar thermochemical applications}.
\newblock \emph{\bibinfo{journal}{Phys. Rev. Mater.}}
  \textbf{\bibinfo{volume}{2}}, \bibinfo{pages}{095401} (\bibinfo{year}{2018}).

\bibitem{Long2020}
\bibinfo{author}{Long, O.~Y.}, \bibinfo{author}{Sai~Gautam, G.} \&
  \bibinfo{author}{Carter, E.~A.}
\newblock \bibinfo{title}{Evaluating optimal {U} for {3d} transition-metal
  oxides within the {SCAN}+{U} framework}.
\newblock \emph{\bibinfo{journal}{Phys. Rev. Mater.}}
  \textbf{\bibinfo{volume}{4}}, \bibinfo{pages}{045401} (\bibinfo{year}{2020}).

\bibitem{Swathilakshmi2023}
\bibinfo{author}{Swathilakshmi, S.}, \bibinfo{author}{Devi, R.} \&
  \bibinfo{author}{Sai~Gautam, G.}
\newblock \bibinfo{title}{Performance of the {r$^2$SCAN} functional in
  transition metal oxides}.
\newblock \emph{\bibinfo{journal}{J. Chem. Theory Comput.}}
  \textbf{\bibinfo{volume}{19}}, \bibinfo{pages}{4202--4215}
  (\bibinfo{year}{2023}).

\bibitem{Wang2006}
\bibinfo{author}{Wang, L.}, \bibinfo{author}{Maxisch, T.} \&
  \bibinfo{author}{Ceder, G.}
\newblock \bibinfo{title}{Oxidation energies of transition metal oxides within
  the {GGA}+{U} framework}.
\newblock \emph{\bibinfo{journal}{Phys. Rev. B}} \textbf{\bibinfo{volume}{73}},
  \bibinfo{pages}{195107} (\bibinfo{year}{2006}).

\bibitem{Chevrier2010}
\bibinfo{author}{Chevrier, V.~L.}, \bibinfo{author}{Ong, S.~P.},
  \bibinfo{author}{Armiento, R.}, \bibinfo{author}{Chan, M. K.~Y.} \&
  \bibinfo{author}{Ceder, G.}
\newblock \bibinfo{title}{Hybrid density functional calculations of redox
  potentials and formation energies of transition metal compounds}.
\newblock \emph{\bibinfo{journal}{Phys. Rev. B}} \textbf{\bibinfo{volume}{82}},
  \bibinfo{pages}{075122} (\bibinfo{year}{2010}).

\bibitem{Jain2011}
\bibinfo{author}{Jain, A.} \emph{et~al.}
\newblock \bibinfo{title}{Formation enthalpies by mixing {GGA} and {GGA}+{U}
  calculations}.
\newblock \emph{\bibinfo{journal}{Phys. Rev. B}} \textbf{\bibinfo{volume}{84}},
  \bibinfo{pages}{045115} (\bibinfo{year}{2011}).

\bibitem{Anisimov1991}
\bibinfo{author}{Anisimov, V.~I.}, \bibinfo{author}{Zaanen, J.} \&
  \bibinfo{author}{Andersen, O.~K.}
\newblock \bibinfo{title}{Band theory and {Mott} insulators: {Hubbard} {U}
  instead of {Stoner} {I}}.
\newblock \emph{\bibinfo{journal}{Phys. Rev. B}} \textbf{\bibinfo{volume}{44}},
  \bibinfo{pages}{943--954} (\bibinfo{year}{1991}).

\bibitem{Dudarev1998}
\bibinfo{author}{Dudarev, S.~L.}, \bibinfo{author}{Botton, G.~A.},
  \bibinfo{author}{Savrasov, S.~Y.}, \bibinfo{author}{Humphreys, C.~J.} \&
  \bibinfo{author}{Sutton, A.~P.}
\newblock \bibinfo{title}{Electron-energy-loss spectra and the structural
  stability of nickel oxide: An {LSDA}+{U} study}.
\newblock \emph{\bibinfo{journal}{Phys. Rev. B}} \textbf{\bibinfo{volume}{57}},
  \bibinfo{pages}{1505--1509} (\bibinfo{year}{1998}).

\bibitem{Artrith2022}
\bibinfo{author}{Artrith, N.}, \bibinfo{author}{Garrido~Torres, J.~A.},
  \bibinfo{author}{Urban, A.} \& \bibinfo{author}{Hybertsen, M.~S.}
\newblock \bibinfo{title}{Data-driven approach to parameterize {SCAN}+{U} for
  an accurate description of 3d transition metal oxide thermochemistry}.
\newblock \emph{\bibinfo{journal}{Phys. Rev. Mater.}}
  \textbf{\bibinfo{volume}{6}}, \bibinfo{pages}{035003} (\bibinfo{year}{2022}).

\bibitem{Tekliye2024}
\bibinfo{author}{Tekliye, D.~B.} \& \bibinfo{author}{Sai~Gautam, G.}
\newblock \bibinfo{title}{Accuracy of {metaGGA} functionals in describing
  transition metal fluorides}.
\newblock \emph{\bibinfo{journal}{Phys. Rev. Mater.}}
  \textbf{\bibinfo{volume}{8}}, \bibinfo{pages}{093801} (\bibinfo{year}{2024}).

\bibitem{Donovan2025}
\bibinfo{author}{Donovan, B.}, \bibinfo{author}{West, A.} \&
  \bibinfo{author}{Urban, A.}
\newblock \bibinfo{title}{{r$^2$SCAN}+{rVV10}+{U} parameterization for 3d
  transition metal sulfides for thermochemistry}.
\newblock \emph{\bibinfo{journal}{J. Chem. Phys.}}
  \textbf{\bibinfo{volume}{163}}, \bibinfo{pages}{034707}
  (\bibinfo{year}{2025}).

\bibitem{Grindy2013}
\bibinfo{author}{Grindy, S.}, \bibinfo{author}{Meredig, B.},
  \bibinfo{author}{Kirklin, S.}, \bibinfo{author}{Saal, J.~E.} \&
  \bibinfo{author}{Wolverton, C.}
\newblock \bibinfo{title}{Approaching chemical accuracy with density functional
  calculations: Diatomic energy corrections}.
\newblock \emph{\bibinfo{journal}{Phys. Rev. B}} \textbf{\bibinfo{volume}{87}},
  \bibinfo{pages}{075150} (\bibinfo{year}{2013}).

\bibitem{Friedrich2019}
\bibinfo{author}{Friedrich, R.} \emph{et~al.}
\newblock \bibinfo{title}{Coordination corrected ab initio formation
  enthalpies}.
\newblock \emph{\bibinfo{journal}{npj Comput. Mater.}}
  \textbf{\bibinfo{volume}{5}}, \bibinfo{pages}{59} (\bibinfo{year}{2019}).

\bibitem{Aykol2014}
\bibinfo{author}{Aykol, M.} \& \bibinfo{author}{Wolverton, C.}
\newblock \bibinfo{title}{Local environment dependent {GGA}+{U} method for
  accurate thermochemistry of transition metal compounds}.
\newblock \emph{\bibinfo{journal}{Phys. Rev. B}} \textbf{\bibinfo{volume}{90}},
  \bibinfo{pages}{115105} (\bibinfo{year}{2014}).

\bibitem{Gong2022}
\bibinfo{author}{Gong, S.} \emph{et~al.}
\newblock \bibinfo{title}{Calibrating {DFT} formation enthalpy calculations by
  multifidelity machine learning}.
\newblock \emph{\bibinfo{journal}{JACS Au}} \textbf{\bibinfo{volume}{2}},
  \bibinfo{pages}{1964--1977} (\bibinfo{year}{2022}).

\bibitem{Bartel2018}
\bibinfo{author}{Bartel, C.~J.} \emph{et~al.}
\newblock \bibinfo{title}{Physical descriptor for the {Gibbs} energy of
  inorganic crystalline solids and temperature-dependent materials chemistry}.
\newblock \emph{\bibinfo{journal}{Nat. Commun.}} \textbf{\bibinfo{volume}{9}},
  \bibinfo{pages}{4168} (\bibinfo{year}{2018}).

\bibitem{Reents2026}
\bibinfo{author}{Reents, T.}, \bibinfo{author}{Bercx, M.} \&
  \bibinfo{author}{Pizzi, G.}
\newblock \bibinfo{title}{Correcting {DFT} formation energies towards
  experimental accuracy using foundational {MLIPs} and latent-feature
  delta-learning} (\bibinfo{year}{2026}).
\newblock
  \bibinfo{eprint}{{\href{https://arxiv.org/abs/2607.18092}{{arXiv:2607.18092}}}}.

\bibitem{Lany2008}
\bibinfo{author}{Lany, S.}
\newblock \bibinfo{title}{Semiconductor thermochemistry in density functional
  calculations}.
\newblock \emph{\bibinfo{journal}{Phys. Rev. B}} \textbf{\bibinfo{volume}{78}},
  \bibinfo{pages}{245207} (\bibinfo{year}{2008}).

\bibitem{Stevanovic2012}
\bibinfo{author}{Stevanovi{\'c}, V.}, \bibinfo{author}{Lany, S.},
  \bibinfo{author}{Zhang, X.} \& \bibinfo{author}{Zunger, A.}
\newblock \bibinfo{title}{Correcting density functional theory for accurate
  predictions of compound enthalpies of formation: Fitted elemental-phase
  reference energies}.
\newblock \emph{\bibinfo{journal}{Phys. Rev. B}} \textbf{\bibinfo{volume}{85}},
  \bibinfo{pages}{115104} (\bibinfo{year}{2012}).

\bibitem{Wang2021}
\bibinfo{author}{Wang, A.} \emph{et~al.}
\newblock \bibinfo{title}{A framework for quantifying uncertainty in {DFT}
  energy corrections}.
\newblock \emph{\bibinfo{journal}{Sci. Rep.}} \textbf{\bibinfo{volume}{11}},
  \bibinfo{pages}{15496} (\bibinfo{year}{2021}).

\bibitem{Cococcioni2005}
\bibinfo{author}{Cococcioni, M.} \& \bibinfo{author}{de~Gironcoli, S.}
\newblock \bibinfo{title}{Linear response approach to the calculation of the
  effective interaction parameters in the {LDA}+{U} method}.
\newblock \emph{\bibinfo{journal}{Phys. Rev. B}} \textbf{\bibinfo{volume}{71}},
  \bibinfo{pages}{035105} (\bibinfo{year}{2005}).

\bibitem{Kulik2011}
\bibinfo{author}{Kulik, H.~J.} \& \bibinfo{author}{Marzari, N.}
\newblock \bibinfo{title}{Accurate potential energy surfaces with a
  {DFT}+{$U(\mathbf{R})$} approach}.
\newblock \emph{\bibinfo{journal}{J. Chem. Phys.}}
  \textbf{\bibinfo{volume}{135}}, \bibinfo{pages}{194105}
  (\bibinfo{year}{2011}).

\bibitem{Moore2024}
\bibinfo{author}{Moore, G.~C.} \emph{et~al.}
\newblock \bibinfo{title}{High-throughput determination of {Hubbard} {U} and
  {Hund} {J} values for transition metal oxides via the linear response
  formalism}.
\newblock \emph{\bibinfo{journal}{Phys. Rev. Mater.}}
  \textbf{\bibinfo{volume}{8}}, \bibinfo{pages}{014409} (\bibinfo{year}{2024}).

\bibitem{Bastonero2025}
\bibinfo{author}{Bastonero, L.} \emph{et~al.}
\newblock \bibinfo{title}{First-principles {Hubbard} parameters with automated
  and reproducible workflows}.
\newblock \emph{\bibinfo{journal}{npj Comput. Mater.}}
  \textbf{\bibinfo{volume}{11}}, \bibinfo{pages}{183} (\bibinfo{year}{2025}).

\bibitem{Aryasetiawan2004}
\bibinfo{author}{Aryasetiawan, F.} \emph{et~al.}
\newblock \bibinfo{title}{Frequency-dependent local interactions and low-energy
  effective models from electronic structure calculations}.
\newblock \emph{\bibinfo{journal}{Phys. Rev. B}} \textbf{\bibinfo{volume}{70}},
  \bibinfo{pages}{195104} (\bibinfo{year}{2004}).

\bibitem{Yu2020}
\bibinfo{author}{Yu, M.}, \bibinfo{author}{Yang, S.}, \bibinfo{author}{Wu, C.}
  \& \bibinfo{author}{Marom, N.}
\newblock \bibinfo{title}{Machine learning the {Hubbard} {U} parameter in
  {DFT}+{U} using {Bayesian} optimization}.
\newblock \emph{\bibinfo{journal}{npj Comput. Mater.}}
  \textbf{\bibinfo{volume}{6}}, \bibinfo{pages}{180} (\bibinfo{year}{2020}).

\bibitem{Cai2024}
\bibinfo{author}{Cai, G.} \emph{et~al.}
\newblock \bibinfo{title}{Predicting structure-dependent {Hubbard} {U}
  parameters via machine learning}.
\newblock \emph{\bibinfo{journal}{Mater. Futures}}
  \textbf{\bibinfo{volume}{3}}, \bibinfo{pages}{025601} (\bibinfo{year}{2024}).

\bibitem{Tavadze2021}
\bibinfo{author}{Tavadze, P.} \emph{et~al.}
\newblock \bibinfo{title}{Exploring {DFT}+{U} parameter space with a {Bayesian}
  calibration assisted by {Markov} chain {Monte} {Carlo} sampling}.
\newblock \emph{\bibinfo{journal}{npj Comput. Mater.}}
  \textbf{\bibinfo{volume}{7}}, \bibinfo{pages}{182} (\bibinfo{year}{2021}).

\bibitem{Xia2024}
\bibinfo{author}{Xia, W.} \emph{et~al.}
\newblock \bibinfo{title}{Machine learning model for the prediction of
  {Hubbard} {U} parameters and its application to {Fe}--{O} systems}.
\newblock \emph{\bibinfo{journal}{J. Chem. Theory Comput.}}
  \textbf{\bibinfo{volume}{20}}, \bibinfo{pages}{10095--10102}
  (\bibinfo{year}{2024}).

\bibitem{Cao2026}
\bibinfo{author}{Cao, B.}, \bibinfo{author}{Xia, W.}, \bibinfo{author}{Zhao,
  J.} \& \bibinfo{author}{Wang, X.}
\newblock \bibinfo{title}{Machine learning-driven {Hubbard} {U} prediction for
  high-throughput study of high-pressure manganese oxides}.
\newblock \emph{\bibinfo{journal}{Phys. Rev. B}}
  \textbf{\bibinfo{volume}{114}}, \bibinfo{pages}{134102}
  (\bibinfo{year}{2026}).

\bibitem{Uhrin2025}
\bibinfo{author}{Uhrin, M.}, \bibinfo{author}{Zadoks, A.},
  \bibinfo{author}{Binci, L.}, \bibinfo{author}{Marzari, N.} \&
  \bibinfo{author}{Timrov, I.}
\newblock \bibinfo{title}{Machine learning {Hubbard} parameters with
  equivariant neural networks}.
\newblock \emph{\bibinfo{journal}{npj Comput. Mater.}}
  \textbf{\bibinfo{volume}{11}}, \bibinfo{pages}{19} (\bibinfo{year}{2025}).

\bibitem{Warford2026}
\bibinfo{author}{Warford, T.}, \bibinfo{author}{Thiemann, F.~L.} \&
  \bibinfo{author}{Cs{\'a}nyi, G.}
\newblock \bibinfo{title}{Better without {U}: impact of selective {Hubbard} {U}
  correction on foundational {MLIPs}}.
\newblock \emph{\bibinfo{journal}{Mach. Learn.: Sci. Technol.}}
  \textbf{\bibinfo{volume}{7}}, \bibinfo{pages}{035033} (\bibinfo{year}{2026}).

\bibitem{Oganov2019}
\bibinfo{author}{Oganov, A.~R.}, \bibinfo{author}{Pickard, C.~J.},
  \bibinfo{author}{Zhu, Q.} \& \bibinfo{author}{Needs, R.~J.}
\newblock \bibinfo{title}{Structure prediction drives materials discovery}.
\newblock \emph{\bibinfo{journal}{Nat. Rev. Mater.}}
  \textbf{\bibinfo{volume}{4}}, \bibinfo{pages}{331--348}
  (\bibinfo{year}{2019}).

\bibitem{Merchant2023}
\bibinfo{author}{Merchant, A.} \emph{et~al.}
\newblock \bibinfo{title}{Scaling deep learning for materials discovery}.
\newblock \emph{\bibinfo{journal}{Nature}} \textbf{\bibinfo{volume}{624}},
  \bibinfo{pages}{80--85} (\bibinfo{year}{2023}).

\bibitem{Horton2025}
\bibinfo{author}{Horton, M.~K.} \emph{et~al.}
\newblock \bibinfo{title}{Accelerated data-driven materials science with the
  {Materials Project}}.
\newblock \emph{\bibinfo{journal}{Nat. Mater.}} \textbf{\bibinfo{volume}{24}},
  \bibinfo{pages}{1522--1532} (\bibinfo{year}{2025}).

\bibitem{Murphy2012}
\bibinfo{author}{Murphy, K.~P.}
\newblock \emph{\bibinfo{title}{Machine Learning: A Probabilistic Perspective}}
   (\bibinfo{publisher}{MIT Press}, \bibinfo{address}{Cambridge, MA},
  \bibinfo{year}{2012}).

\bibitem{Popov2021}
\bibinfo{author}{Popov, M.~N.}, \bibinfo{author}{Spitaler, J.},
  \bibinfo{author}{Romaner, L.}, \bibinfo{author}{Bedoya-Mart{\'i}nez, N.} \&
  \bibinfo{author}{Hammer, R.}
\newblock \bibinfo{title}{Bayesian optimization of {Hubbard} {U}'s for
  investigating {InGaN} superlattices}.
\newblock \emph{\bibinfo{journal}{Electron. Mater.}}
  \textbf{\bibinfo{volume}{2}}, \bibinfo{pages}{370--381}
  (\bibinfo{year}{2021}).

\bibitem{Yu2023}
\bibinfo{author}{Yu, W.} \emph{et~al.}
\newblock \bibinfo{title}{Active learning the high-dimensional transferable
  {Hubbard} {U} and {V} parameters in the {DFT} + {U} + {V} scheme}.
\newblock \emph{\bibinfo{journal}{J. Chem. Theory Comput.}}
  \textbf{\bibinfo{volume}{19}}, \bibinfo{pages}{6425--6433}
  (\bibinfo{year}{2023}).

\bibitem{Zhuo2018}
\bibinfo{author}{Zhuo, Y.}, \bibinfo{author}{Mansouri~Tehrani, A.} \&
  \bibinfo{author}{Brgoch, J.}
\newblock \bibinfo{title}{Predicting the band gaps of inorganic solids by
  machine learning}.
\newblock \emph{\bibinfo{journal}{J. Phys. Chem. Lett.}}
  \textbf{\bibinfo{volume}{9}}, \bibinfo{pages}{1668--1673}
  (\bibinfo{year}{2018}).

\bibitem{Kiselyova2016}
\bibinfo{author}{Kiselyova, N.~N.}, \bibinfo{author}{Dudarev, V.~A.} \&
  \bibinfo{author}{Korzhuyev, M.~A.}
\newblock \bibinfo{title}{Database on the bandgap of inorganic substances and
  materials}.
\newblock \emph{\bibinfo{journal}{Inorg. Mater. Appl. Res.}}
  \textbf{\bibinfo{volume}{7}}, \bibinfo{pages}{34--39} (\bibinfo{year}{2016}).

\bibitem{Strehlow1973}
\bibinfo{author}{Strehlow, W.~H.} \& \bibinfo{author}{Cook, E.~L.}
\newblock \bibinfo{title}{Compilation of energy band gaps in elemental and
  binary compound semiconductors and insulators}.
\newblock \emph{\bibinfo{journal}{J. Phys. Chem. Ref. Data}}
  \textbf{\bibinfo{volume}{2}}, \bibinfo{pages}{163--200}
  (\bibinfo{year}{1973}).

\bibitem{Joshi1990}
\bibinfo{author}{Joshi, N.~V.}
\newblock \emph{\bibinfo{title}{Photoconductivity: Art, Science, and
  Technology}}  (\bibinfo{publisher}{Marcel Dekker}, \bibinfo{address}{New
  York}, \bibinfo{year}{1990}).

\bibitem{Madelung2004}
\bibinfo{author}{Madelung, O.}
\newblock \emph{\bibinfo{title}{Semiconductors: Data Handbook}}
  (\bibinfo{publisher}{Springer}, \bibinfo{address}{New York},
  \bibinfo{year}{2004}).

\bibitem{Dunn2020}
\bibinfo{author}{Dunn, A.}, \bibinfo{author}{Wang, Q.},
  \bibinfo{author}{Ganose, A.}, \bibinfo{author}{Dopp, D.} \&
  \bibinfo{author}{Jain, A.}
\newblock \bibinfo{title}{Benchmarking materials property prediction methods:
  the {Matbench} test set and {Automatminer} reference algorithm}.
\newblock \emph{\bibinfo{journal}{npj Comput. Mater.}}
  \textbf{\bibinfo{volume}{6}}, \bibinfo{pages}{138} (\bibinfo{year}{2020}).

\bibitem{Hautier2012}
\bibinfo{author}{Hautier, G.}, \bibinfo{author}{Ong, S.~P.},
  \bibinfo{author}{Jain, A.}, \bibinfo{author}{Moore, C.~J.} \&
  \bibinfo{author}{Ceder, G.}
\newblock \bibinfo{title}{Accuracy of density functional theory in predicting
  formation energies of ternary oxides from binary oxides and its implication
  on phase stability}.
\newblock \emph{\bibinfo{journal}{Phys. Rev. B}} \textbf{\bibinfo{volume}{85}},
  \bibinfo{pages}{155208} (\bibinfo{year}{2012}).

\bibitem{Blochl1994}
\bibinfo{author}{Bl{\"o}chl, P.~E.}
\newblock \bibinfo{title}{Projector augmented-wave method}.
\newblock \emph{\bibinfo{journal}{Phys. Rev. B}} \textbf{\bibinfo{volume}{50}},
  \bibinfo{pages}{17953--17979} (\bibinfo{year}{1994}).

\bibitem{Kresse1999}
\bibinfo{author}{Kresse, G.} \& \bibinfo{author}{Joubert, D.}
\newblock \bibinfo{title}{From ultrasoft pseudopotentials to the projector
  augmented-wave method}.
\newblock \emph{\bibinfo{journal}{Phys. Rev. B}} \textbf{\bibinfo{volume}{59}},
  \bibinfo{pages}{1758--1775} (\bibinfo{year}{1999}).

\bibitem{Kresse1993}
\bibinfo{author}{Kresse, G.} \& \bibinfo{author}{Hafner, J.}
\newblock \bibinfo{title}{Ab initio molecular dynamics for liquid metals}.
\newblock \emph{\bibinfo{journal}{Phys. Rev. B}} \textbf{\bibinfo{volume}{47}},
  \bibinfo{pages}{558--561} (\bibinfo{year}{1993}).

\bibitem{Kresse1996a}
\bibinfo{author}{Kresse, G.} \& \bibinfo{author}{Furthm{\"u}ller, J.}
\newblock \bibinfo{title}{Efficiency of ab-initio total energy calculations for
  metals and semiconductors using a plane-wave basis set}.
\newblock \emph{\bibinfo{journal}{Comput. Mater. Sci.}}
  \textbf{\bibinfo{volume}{6}}, \bibinfo{pages}{15--50} (\bibinfo{year}{1996}).

\bibitem{Kresse1996b}
\bibinfo{author}{Kresse, G.} \& \bibinfo{author}{Furthm{\"u}ller, J.}
\newblock \bibinfo{title}{Efficient iterative schemes for ab initio
  total-energy calculations using a plane-wave basis set}.
\newblock \emph{\bibinfo{journal}{Phys. Rev. B}} \textbf{\bibinfo{volume}{54}},
  \bibinfo{pages}{11169--11186} (\bibinfo{year}{1996}).

\bibitem{Rumble2025CRC}
\bibinfo{editor}{Rumble, J.~R.}, \bibinfo{editor}{Bruno, T.~J.},
  \bibinfo{editor}{Burgess, D.~R., Jr.} \& \bibinfo{editor}{McEwen, L.~R.}
  (eds) \emph{\bibinfo{title}{{CRC Handbook of Chemistry and Physics: A
  Ready-Reference Book of Chemical and Physical Data}}}
  \bibinfo{edition}{106th} edn (\bibinfo{publisher}{CRC Press},
  \bibinfo{address}{Boca Raton, London, New York}, \bibinfo{year}{2025}).

\bibitem{Blundell2001}
\bibinfo{author}{Blundell, S.}
\newblock \emph{\bibinfo{title}{Magnetism in Condensed Matter}} Oxford Master
  Series in Condensed Matter Physics (\bibinfo{publisher}{Oxford University
  Press}, \bibinfo{address}{Oxford}, \bibinfo{year}{2001}).

\bibitem{PerezMato2012}
\bibinfo{author}{Perez-Mato, J.~M.}, \bibinfo{author}{Ribeiro, J.~L.},
  \bibinfo{author}{Petricek, V.} \& \bibinfo{author}{Aroyo, M.~I.}
\newblock \bibinfo{title}{Magnetic superspace groups and symmetry constraints
  in incommensurate magnetic phases}.
\newblock \emph{\bibinfo{journal}{J. Phys.: Condens. Matter}}
  \textbf{\bibinfo{volume}{24}}, \bibinfo{pages}{163201}
  (\bibinfo{year}{2012}).

\bibitem{Lawson1994}
\bibinfo{author}{Lawson, A.~C.} \emph{et~al.}
\newblock \bibinfo{title}{Magnetic and crystallographic order in
  $\alpha$-manganese}.
\newblock \emph{\bibinfo{journal}{J. Appl. Phys.}}
  \textbf{\bibinfo{volume}{76}}, \bibinfo{pages}{7049} (\bibinfo{year}{1994}).

\bibitem{Robie1995}
\bibinfo{author}{Robie, R.~A.} \& \bibinfo{author}{Hemingway, B.~S.}
\newblock \bibinfo{title}{Thermodynamic properties of minerals and related
  substances at 298.15 {K} and 1 bar (10$^5$ pascals) pressure and at higher
  temperatures}.
\newblock \bibinfo{type}{Bulletin} \bibinfo{number}{2131},
  \bibinfo{institution}{U.S. Geological Survey} (\bibinfo{year}{1995}).

\bibitem{Yungman1999}
\bibinfo{editor}{Yungman, V.~S.} (ed.) \emph{\bibinfo{title}{Thermal Constants
  of Substances}}  (\bibinfo{publisher}{Wiley}, \bibinfo{address}{New York},
  \bibinfo{year}{1999}).
\newblock
  \urlprefix\url{https://www.chem.msu.ru/cgi-bin/tkv.pl?show=welcome.html}.

\bibitem{Cox1989}
\bibinfo{author}{Cox, J.~D.}, \bibinfo{author}{Wagman, D.~D.} \&
  \bibinfo{author}{Medvedev, V.~A.}
\newblock \emph{\bibinfo{title}{{CODATA} Key Values for Thermodynamics}}
  (\bibinfo{publisher}{Hemisphere}, \bibinfo{address}{New York},
  \bibinfo{year}{1989}).
\newblock
  \urlprefix\url{https://www.codata.info/resources/databases/key1.html}.

\bibitem{Chase1998}
\bibinfo{author}{Chase, M.~W., Jr.}
\newblock \emph{\bibinfo{title}{{NIST-JANAF} Thermochemical Tables}}
  \bibinfo{edition}{4th} edn.
\newblock No.~\bibinfo{number}{9} in \bibinfo{series}{J. Phys. Chem. Ref. Data
  Monogr.} (\bibinfo{publisher}{American Institute of Physics},
  \bibinfo{year}{1998}).

\bibitem{Lamoreaux1984}
\bibinfo{author}{Lamoreaux, R.~H.} \& \bibinfo{author}{Hildenbrand, D.~L.}
\newblock \bibinfo{title}{High temperature vaporization behavior of oxides.
  {I}. alkali metal binary oxides}.
\newblock \emph{\bibinfo{journal}{J. Phys. Chem. Ref. Data}}
  \textbf{\bibinfo{volume}{13}}, \bibinfo{pages}{151--173}
  (\bibinfo{year}{1984}).

\bibitem{Parker1971}
\bibinfo{author}{Parker, V.~B.}, \bibinfo{author}{Wagman, D.~D.} \&
  \bibinfo{author}{Evans, W.~H.}
\newblock \bibinfo{title}{Selected values of chemical thermodynamic properties:
  Tables for the alkaline earth elements (elements 92 through 97 in the
  standard order of arrangement)}.
\newblock \bibinfo{type}{Technical Note} \bibinfo{number}{270-6},
  \bibinfo{institution}{National Bureau of Standards} (\bibinfo{year}{1971}).

\bibitem{Huntelaar1992Ba}
\bibinfo{author}{Huntelaar, M.~E.}, \bibinfo{author}{Cordfunke, E. H.~P.} \&
  \bibinfo{author}{Ouweltjes, W.}
\newblock \bibinfo{title}{The standard molar enthalpies of formation of
  {BaSiO$_3$(s)} and {Ba$_2$SiO$_4$(s)}}.
\newblock \emph{\bibinfo{journal}{J. Chem. Thermodyn.}}
  \textbf{\bibinfo{volume}{24}}, \bibinfo{pages}{1099--1102}
  (\bibinfo{year}{1992}).

\bibitem{Huntelaar1992Sr}
\bibinfo{author}{Huntelaar, M.~E.}, \bibinfo{author}{Cordfunke, E. H.~P.} \&
  \bibinfo{author}{Ouweltjes, W.}
\newblock \bibinfo{title}{The standard molar enthalpies of formation of
  {SrSiO$_3$(s)} and {Sr$_2$SiO$_4$(s)}}.
\newblock \emph{\bibinfo{journal}{J. Chem. Thermodyn.}}
  \textbf{\bibinfo{volume}{24}}, \bibinfo{pages}{139--143}
  (\bibinfo{year}{1992}).

\bibitem{Wagman1982}
\bibinfo{author}{Wagman, D.~D.} \emph{et~al.}
\newblock \bibinfo{title}{The {NBS} tables of chemical thermodynamic
  properties: Selected values for inorganic and {C$_1$} and {C$_2$} organic
  substances in {SI} units}.
\newblock \emph{\bibinfo{journal}{J. Phys. Chem. Ref. Data}}
  \textbf{\bibinfo{volume}{11}}, \bibinfo{pages}{1--392}
  (\bibinfo{year}{1982}).

\bibitem{Reed2020}
\bibinfo{author}{Reed, J.~J.}
\newblock \bibinfo{title}{The {NBS} tables of chemical thermodynamic
  properties: Selected values for inorganic and {C$_1$} and {C$_2$} organic
  substances in {SI} units}.
\newblock \bibinfo{howpublished}{{NIST} Public Data Repository}
  (\bibinfo{year}{2020}).

\bibitem{Kubaschewski1993}
\bibinfo{author}{Kubaschewski, O.}, \bibinfo{author}{Alcock, C.~B.} \&
  \bibinfo{author}{Spencer, P.~J.}
\newblock \emph{\bibinfo{title}{Materials Thermochemistry}}
  \bibinfo{edition}{6th} edn (\bibinfo{publisher}{Pergamon Press},
  \bibinfo{address}{Oxford}, \bibinfo{year}{1993}).

\bibitem{LandoltBornstein}
\bibinfo{title}{{Landolt-B{\"o}rnstein} database}.
\newblock \bibinfo{howpublished}{Springer-Verlag, Berlin Heidelberg; Material
  Phases Data System (MPDS), Switzerland; National Institute for Materials
  Science (NIMS), Japan}.

\bibitem{Wagman1969}
\bibinfo{author}{Wagman, D.~D.} \emph{et~al.}
\newblock \bibinfo{title}{Selected values of chemical thermodynamic properties:
  Tables for elements 35 through 53 in the standard order of arrangement}.
\newblock \bibinfo{type}{Technical Note} \bibinfo{number}{270-4},
  \bibinfo{institution}{National Bureau of Standards} (\bibinfo{year}{1969}).

\bibitem{Wagman1981}
\bibinfo{author}{Wagman, D.~D.}, \bibinfo{author}{Evans, W.~H.},
  \bibinfo{author}{Parker, V.~B.}, \bibinfo{author}{Schumm, R.~H.} \&
  \bibinfo{author}{Nuttall, R.~L.}
\newblock \bibinfo{title}{Selected values of chemical thermodynamic properties:
  Compounds of uranium, protactinium, thorium, actinium, and the alkali
  metals}.
\newblock \bibinfo{type}{Technical Note} \bibinfo{number}{270-8},
  \bibinfo{institution}{National Bureau of Standards} (\bibinfo{year}{1981}).

\bibitem{Lemire2013}
\bibinfo{author}{Lemire, R.~J.} \emph{et~al.}
\newblock \emph{\bibinfo{title}{Chemical Thermodynamics of Iron, Part 1}} Vol.
  \bibinfo{volume}{13a} of \emph{\bibinfo{series}{Chemical Thermodynamics}}
  (\bibinfo{publisher}{OECD Publishing}, \bibinfo{address}{Paris},
  \bibinfo{year}{2013}).

\bibitem{Rard1999}
\bibinfo{author}{Rard, J.~A.}, \bibinfo{author}{Rand, M.~H.},
  \bibinfo{author}{Anderegg, G.} \& \bibinfo{author}{Wanner, H.}
\newblock \emph{\bibinfo{title}{Chemical Thermodynamics of Technetium}}
  Vol.~\bibinfo{volume}{3} of \emph{\bibinfo{series}{Chemical Thermodynamics}}
  (\bibinfo{publisher}{Elsevier}, \bibinfo{address}{Amsterdam},
  \bibinfo{year}{1999}).

\bibitem{ONeill1997}
\bibinfo{author}{O'Neill, H. S.~C.} \& \bibinfo{author}{Nell, J.}
\newblock \bibinfo{title}{Gibbs free energies of formation of {RuO$_2$},
  {IrO$_2$}, and {OsO$_2$}: A high-temperature electrochemical and calorimetric
  study}.
\newblock \emph{\bibinfo{journal}{Geochim. Cosmochim. Acta}}
  \textbf{\bibinfo{volume}{61}}, \bibinfo{pages}{5279--5293}
  (\bibinfo{year}{1997}).

\bibitem{Chatillon2018}
\bibinfo{author}{Chatillon, C.}, \bibinfo{author}{Nuta, I.},
  \bibinfo{author}{Roki, F.-Z.} \& \bibinfo{author}{Fischer, E.}
\newblock \bibinfo{title}{Chemical thermodynamics of {RuO$_2$(s)}}.
\newblock \emph{\bibinfo{journal}{J. Nucl. Mater.}}
  \textbf{\bibinfo{volume}{509}}, \bibinfo{pages}{742--751}
  (\bibinfo{year}{2018}).

\bibitem{Arjmand2015}
\bibinfo{author}{Arjmand, M.}, \bibinfo{author}{Knee, C.~S.},
  \bibinfo{author}{Leion, H.} \& \bibinfo{author}{Mattisson, T.}
\newblock \bibinfo{title}{Standard enthalpy of formation of {CuAl$_2$O$_4$}
  revisited}.
\newblock \emph{\bibinfo{journal}{Chem. Eng. Commun.}}
  \textbf{\bibinfo{volume}{202}}, \bibinfo{pages}{694--697}
  (\bibinfo{year}{2015}).

\bibitem{Wang2004LiMO2}
\bibinfo{author}{Wang, M.} \& \bibinfo{author}{Navrotsky, A.}
\newblock \bibinfo{title}{Enthalpy of formation of {LiNiO$_2$}, {LiCoO$_2$} and
  their solid solution, {LiNi$_{1-x}$Co$_x$O$_2$}}.
\newblock \emph{\bibinfo{journal}{Solid State Ion.}}
  \textbf{\bibinfo{volume}{166}}, \bibinfo{pages}{167--173}
  (\bibinfo{year}{2004}).

\bibitem{Wang2005LiMnO2}
\bibinfo{author}{Wang, M.} \& \bibinfo{author}{Navrotsky, A.}
\newblock \bibinfo{title}{{LiMO$_2$} ({M} = {Mn}, {Fe}, and {Co}): Energetics,
  polymorphism and phase transformation}.
\newblock \emph{\bibinfo{journal}{J. Solid State Chem.}}
  \textbf{\bibinfo{volume}{178}}, \bibinfo{pages}{1230--1240}
  (\bibinfo{year}{2005}).

\bibitem{Cheng2005}
\bibinfo{author}{Cheng, J.}, \bibinfo{author}{Navrotsky, A.},
  \bibinfo{author}{Zhou, X.-D.} \& \bibinfo{author}{Anderson, H.~U.}
\newblock \bibinfo{title}{Enthalpies of formation of {LaMO$_3$} perovskites
  ({M} = {Cr}, {Fe}, {Co}, and {Ni})}.
\newblock \emph{\bibinfo{journal}{J. Mater. Res.}}
  \textbf{\bibinfo{volume}{20}}, \bibinfo{pages}{191--200}
  (\bibinfo{year}{2005}).

\bibitem{Whittingham1976}
\bibinfo{author}{Whittingham, M.~S.}
\newblock \bibinfo{title}{Electrical energy storage and intercalation
  chemistry}.
\newblock \emph{\bibinfo{journal}{Science}} \textbf{\bibinfo{volume}{192}},
  \bibinfo{pages}{1126--1127} (\bibinfo{year}{1976}).

\bibitem{Reimers1992}
\bibinfo{author}{Reimers, J.~N.} \& \bibinfo{author}{Dahn, J.~R.}
\newblock \bibinfo{title}{Electrochemical and in situ {X}-ray diffraction
  studies of lithium intercalation in {Li$_x$CoO$_2$}}.
\newblock \emph{\bibinfo{journal}{J. Electrochem. Soc.}}
  \textbf{\bibinfo{volume}{139}}, \bibinfo{pages}{2091--2097}
  (\bibinfo{year}{1992}).

\bibitem{Delmas1994}
\bibinfo{author}{Delmas, C.}, \bibinfo{author}{Cognac-Auradou, H.},
  \bibinfo{author}{Cocciantelli, J.~M.}, \bibinfo{author}{M{\'e}n{\'e}trier,
  M.} \& \bibinfo{author}{Doumerc, J.~P.}
\newblock \bibinfo{title}{The {Li$_x$V$_2$O$_5$} system: An overview of the
  structure modifications induced by the lithium intercalation}.
\newblock \emph{\bibinfo{journal}{Solid State Ion.}}
  \textbf{\bibinfo{volume}{69}}, \bibinfo{pages}{257--264}
  (\bibinfo{year}{1994}).

\bibitem{Colbow1989}
\bibinfo{author}{Colbow, K.~M.}, \bibinfo{author}{Dahn, J.~R.} \&
  \bibinfo{author}{Haering, R.~R.}
\newblock \bibinfo{title}{Structure and electrochemistry of the spinel oxides
  {LiTi$_2$O$_4$} and {Li$_{4/3}$Ti$_{5/3}$O$_4$}}.
\newblock \emph{\bibinfo{journal}{J. Power Sources}}
  \textbf{\bibinfo{volume}{26}}, \bibinfo{pages}{397--402}
  (\bibinfo{year}{1989}).

\bibitem{Zagorac2019}
\bibinfo{author}{Zagorac, D.}, \bibinfo{author}{M{\"u}ller, H.},
  \bibinfo{author}{Ruehl, S.}, \bibinfo{author}{Zagorac, J.} \&
  \bibinfo{author}{Rehme, S.}
\newblock \bibinfo{title}{Recent developments in the {Inorganic Crystal
  Structure Database}: theoretical crystal structure data and related
  features}.
\newblock \emph{\bibinfo{journal}{J. Appl. Crystallogr.}}
  \textbf{\bibinfo{volume}{52}}, \bibinfo{pages}{918--925}
  (\bibinfo{year}{2019}).

\bibitem{Delmas1980}
\bibinfo{author}{Delmas, C.}, \bibinfo{author}{Fouassier, C.} \&
  \bibinfo{author}{Hagenmuller, P.}
\newblock \bibinfo{title}{Structural classification and properties of the
  layered oxides}.
\newblock \emph{\bibinfo{journal}{Physica B+C}} \textbf{\bibinfo{volume}{99}},
  \bibinfo{pages}{81--85} (\bibinfo{year}{1980}).

\bibitem{Ong2013}
\bibinfo{author}{Ong, S.~P.} \emph{et~al.}
\newblock \bibinfo{title}{Python materials genomics (pymatgen): A robust,
  open-source python library for materials analysis}.
\newblock \emph{\bibinfo{journal}{Comput. Mater. Sci.}}
  \textbf{\bibinfo{volume}{68}}, \bibinfo{pages}{314--319}
  (\bibinfo{year}{2013}).

\bibitem{George2026}
\bibinfo{author}{George, J.} \emph{et~al.}
\newblock \bibinfo{title}{pymatgen: 15 years of community growth, new
  functionality, and future prospects} (\bibinfo{year}{2026}).
\newblock
  \bibinfo{eprint}{{\href{https://arxiv.org/abs/2610.23538}{{arXiv:2610.23538}}}}.

\bibitem{Horton2019}
\bibinfo{author}{Horton, M.~K.}, \bibinfo{author}{Montoya, J.~H.},
  \bibinfo{author}{Liu, M.} \& \bibinfo{author}{Persson, K.~A.}
\newblock \bibinfo{title}{High-throughput prediction of the ground-state
  collinear magnetic order of inorganic materials using density functional
  theory}.
\newblock \emph{\bibinfo{journal}{npj Comput. Mater.}}
  \textbf{\bibinfo{volume}{5}}, \bibinfo{pages}{64} (\bibinfo{year}{2019}).

\bibitem{Hart2008}
\bibinfo{author}{Hart, G. L.~W.} \& \bibinfo{author}{Forcade, R.~W.}
\newblock \bibinfo{title}{Algorithm for generating derivative structures}.
\newblock \emph{\bibinfo{journal}{Phys. Rev. B}} \textbf{\bibinfo{volume}{77}},
  \bibinfo{pages}{224115} (\bibinfo{year}{2008}).

\bibitem{Hart2012}
\bibinfo{author}{Hart, G. L.~W.}, \bibinfo{author}{Nelson, L.~J.} \&
  \bibinfo{author}{Forcade, R.~W.}
\newblock \bibinfo{title}{Generating derivative structures at a fixed
  concentration}.
\newblock \emph{\bibinfo{journal}{Comput. Mater. Sci.}}
  \textbf{\bibinfo{volume}{59}}, \bibinfo{pages}{101--107}
  (\bibinfo{year}{2012}).

\bibitem{Fritsch1984}
\bibinfo{author}{Fritsch, F.~N.} \& \bibinfo{author}{Butland, J.}
\newblock \bibinfo{title}{A method for constructing local monotone piecewise
  cubic interpolants}.
\newblock \emph{\bibinfo{journal}{SIAM J. Sci. Stat. Comput.}}
  \textbf{\bibinfo{volume}{5}}, \bibinfo{pages}{300--304}
  (\bibinfo{year}{1984}).

\bibitem{Virtanen2020}
\bibinfo{author}{Virtanen, P.} \emph{et~al.}
\newblock \bibinfo{title}{{SciPy} 1.0: fundamental algorithms for scientific
  computing in {Python}}.
\newblock \emph{\bibinfo{journal}{Nat. Methods}} \textbf{\bibinfo{volume}{17}},
  \bibinfo{pages}{261--272} (\bibinfo{year}{2020}).

\bibitem{terBraak2008}
\bibinfo{author}{ter Braak, C. J.~F.} \& \bibinfo{author}{Vrugt, J.~A.}
\newblock \bibinfo{title}{Differential {Evolution} {Markov} chain with snooker
  updater and fewer chains}.
\newblock \emph{\bibinfo{journal}{Stat. Comput.}}
  \textbf{\bibinfo{volume}{18}}, \bibinfo{pages}{435--446}
  (\bibinfo{year}{2008}).

\bibitem{AbrilPla2023}
\bibinfo{author}{Abril-Pla, O.} \emph{et~al.}
\newblock \bibinfo{title}{{PyMC}: a modern, and comprehensive probabilistic
  programming framework in {Python}}.
\newblock \emph{\bibinfo{journal}{PeerJ Comput. Sci.}}
  \textbf{\bibinfo{volume}{9}}, \bibinfo{pages}{e1516} (\bibinfo{year}{2023}).

\bibitem{Vehtari2021}
\bibinfo{author}{Vehtari, A.}, \bibinfo{author}{Gelman, A.},
  \bibinfo{author}{Simpson, D.}, \bibinfo{author}{Carpenter, B.} \&
  \bibinfo{author}{B{\"u}rkner, P.-C.}
\newblock \bibinfo{title}{Rank-normalization, folding, and localization: an
  improved {$\widehat{R}$} for assessing convergence of {MCMC} (with
  discussion)}.
\newblock \emph{\bibinfo{journal}{Bayesian Anal.}}
  \textbf{\bibinfo{volume}{16}}, \bibinfo{pages}{667--718}
  (\bibinfo{year}{2021}).

\bibitem{Manthiram1989}
\bibinfo{author}{Manthiram, A.} \& \bibinfo{author}{Goodenough, J.~B.}
\newblock \bibinfo{title}{Lithium insertion into {Fe$_2$(SO$_4$)$_3$}
  frameworks}.
\newblock \emph{\bibinfo{journal}{J. Power Sources}}
  \textbf{\bibinfo{volume}{26}}, \bibinfo{pages}{403--408}
  (\bibinfo{year}{1989}).

\bibitem{Kerr2000}
\bibinfo{author}{Kerr, T.~A.}, \bibinfo{author}{Gaubicher, J.} \&
  \bibinfo{author}{Nazar, L.~F.}
\newblock \bibinfo{title}{Highly reversible {Li} insertion at 4~{V} in
  $\varepsilon$-{VOPO$_4$}/$\alpha$-{LiVOPO$_4$} cathodes}.
\newblock \emph{\bibinfo{journal}{Electrochem. Solid-State Lett.}}
  \textbf{\bibinfo{volume}{3}}, \bibinfo{pages}{460} (\bibinfo{year}{2000}).

\bibitem{Reynaud2012}
\bibinfo{author}{Reynaud, M.} \emph{et~al.}
\newblock \bibinfo{title}{{Li$_2$Fe(SO$_4$)$_2$} as a 3.83~{V} positive
  electrode material}.
\newblock \emph{\bibinfo{journal}{Electrochem. Commun.}}
  \textbf{\bibinfo{volume}{21}}, \bibinfo{pages}{77--80}
  (\bibinfo{year}{2012}).

\bibitem{Delmas1988}
\bibinfo{author}{Delmas, C.}, \bibinfo{author}{Nadiri, A.} \&
  \bibinfo{author}{Soubeyroux, J.~L.}
\newblock \bibinfo{title}{The nasicon-type titanium phosphates
  {ATi$_2$(PO$_4$)$_3$} ({A} = {Li}, {Na}) as electrode materials}.
\newblock \emph{\bibinfo{journal}{Solid State Ion.}}
  \textbf{\bibinfo{volume}{28--30}}, \bibinfo{pages}{419--423}
  (\bibinfo{year}{1988}).

\bibitem{Yin2003}
\bibinfo{author}{Yin, S.-C.}, \bibinfo{author}{Grondey, H.},
  \bibinfo{author}{Strobel, P.}, \bibinfo{author}{Anne, M.} \&
  \bibinfo{author}{Nazar, L.~F.}
\newblock \bibinfo{title}{Electrochemical property: structure relationships in
  monoclinic {Li$_{3-y}$V$_2$(PO$_4$)$_3$}}.
\newblock \emph{\bibinfo{journal}{J. Am. Chem. Soc.}}
  \textbf{\bibinfo{volume}{125}}, \bibinfo{pages}{10402--10411}
  (\bibinfo{year}{2003}).

\bibitem{Recham2010}
\bibinfo{author}{Recham, N.} \emph{et~al.}
\newblock \bibinfo{title}{A 3.6~{V} lithium-based fluorosulphate insertion
  positive electrode for lithium-ion batteries}.
\newblock \emph{\bibinfo{journal}{Nat. Mater.}} \textbf{\bibinfo{volume}{9}},
  \bibinfo{pages}{68--74} (\bibinfo{year}{2010}).

\bibitem{Barker2003}
\bibinfo{author}{Barker, J.}, \bibinfo{author}{Saidi, M.~Y.} \&
  \bibinfo{author}{Swoyer, J.~L.}
\newblock \bibinfo{title}{Electrochemical insertion properties of the novel
  lithium vanadium fluorophosphate, {LiVPO$_4$F}}.
\newblock \emph{\bibinfo{journal}{J. Electrochem. Soc.}}
  \textbf{\bibinfo{volume}{150}}, \bibinfo{pages}{A1394}
  (\bibinfo{year}{2003}).

\bibitem{Grazulis2012}
\bibinfo{author}{Gra{\v{z}}ulis, S.} \emph{et~al.}
\newblock \bibinfo{title}{Crystallography open database ({COD}): an open-access
  collection of crystal structures and platform for world-wide collaboration}.
\newblock \emph{\bibinfo{journal}{Nucleic Acids Res.}}
  \textbf{\bibinfo{volume}{40}}, \bibinfo{pages}{D420--D427}
  (\bibinfo{year}{2012}).

\bibitem{Ward2018}
\bibinfo{author}{Ward, L.} \emph{et~al.}
\newblock \bibinfo{title}{Matminer: An open source toolkit for materials data
  mining}.
\newblock \emph{\bibinfo{journal}{Comput. Mater. Sci.}}
  \textbf{\bibinfo{volume}{152}}, \bibinfo{pages}{60--69}
  (\bibinfo{year}{2018}).

\end{thebibliography}
\end{document}